%% file: FN-790.tex
\documentclass[11pt]{article}
\usepackage{graphicx,times,latexsym,epsfig,rotate,pslatex}

\input  fnaldocheader
\fnaldocheader{FERMILAB-FN-0790-AD}{July 2006}

\begin{document}

\title{Machine-Related Backgrounds in the SiD Detector at ILC\thanks{
Work supported by the Universities Research Association, Inc., under contract
DE-AC02-76CH03000 with the U.~S.~Department of Energy.}}

\author{D.S.~Denisov, N.V.~Mokhov, S.I.~Striganov\\
\footnotesize \em Fermilab, P.O. Box 500, Batavia, IL 60510\\
M.A.~Kostin\\
\footnotesize \em NSCL, Michigan State University, East Lansing, MI 48824\\
I.S.~Tropin\\
\footnotesize \em Tomsk Polytechnic University, Tomsk, 634034, Russia}

\date{\today}

\maketitle 
\begin{abstract}
With a multi-stage collimation system and magnetic iron spoilers in 
the tunnel, the background particle fluxes on the ILC detector  can be
substantially reduced. At the same time,
beam-halo interactions with collimators and protective masks in the 
beam delivery system create fluxes of muons and other secondary particles
which can still exceed the tolerable levels for some of the ILC sub-detectors.
Results of modeling of such backgrounds in comparison to those from
the $e^+ e^-$ interactions are presented in this paper for the SiD detector.
\end{abstract}

\newpage

\section{Introduction}

The collimators of the International Linear Collider (ILC) Beam Delivery System (BDS)
are intended to localize the beam loss 
in dedicated regions far from the Interaction Point (IP) to substantially
reduce backgrounds in the collider detectors~\cite{BDS}. 
Particle fluxes resulting from the interactions of beam halo with the collimators,
protective masks and other limiting apertures could still exceed 
tolerable levels for some of the ILC sub-detectors. Magnetic spoilers in the tunnel can 
reduce muon fluxes substantially~\cite{snow05}. 
Response of the Silicon Detector (SiD) sub-detectors~\cite{sid} to these backgrounds
is calculated and presented in this report.

\section{BDS and Detector Models, Scraping Rate and Beam Parameters}

Following
the SLAC Linear Collider (SLC) experience, it is assumed that the ILC collimation system
cuts 0.1\% of the beam outside of a predefined beam envelope.
Such scraping rate at the SLC could be explained by absence
of pre-linac collimation and by tails coming from the dumping rings. Using the same loss rate
in the ILC BDS seems to be conservative, but it is accepted at the present stage as a specification
for the collimation system and BDS designs~\cite{BDS,comp03}. Details of the BDS and collimation
system designs and calculated beam loss distributions in the region are described
elsewhere~\cite{BDS,snow-sum,epac06}.
 
Beam losses in the BDS are simulated with the STRUCT code~\cite{struct}.
Starting from beam loss distributions on the betatron and momentum spoilers SP2, SP4, SPEX,
full 3-D shower simulations through
the entire 1.8-km long BDS system are performed with the MARS15 code~\cite{mars}.
A sketch of the ILC collimation system layout used in the  calculations is presented in Fig.~1. The MARS15
model of the BDS includes all the primary collimators (spoilers),
absorbers, protection collimators, synchrotron radiation masks,
focusing and bending magnets with proper materials, 3-D geometry
and magnetic fields, tunnel walls and surrounding dirt~\cite{BDS}. 
These calculations provide particle parameters
and tagging information with the cutoff energy of 0.1~MeV at the entrance to the detector.

The source of the muon component of BDS backgrounds is concentrated in the collimation region 800 to 1500~m
from the IP. Muon flux from this source on the detector can be substantially reduced
by massive magnetic steel blocks in the tunnel closer to the experimental
hall~\cite{keller93}.
With two spoilers 9 and 18-m long at 648 and  331~m from the IP, the muon fluxes
at the detector can be reduced by a few thousand times, as was shown with the MUCARLO and MARS15 
codes~\cite{snow05}.
Each muon spoiler consists of two steel parts with magnetic coils which provide the opposite field
polarities in order to compensate field in the beam pipe. The magnetic field in 
the iron is 1.5~T. The gap between the parts accommodates the beam pipe.
The winding slots in the center of each iron part are 10-cm wide and 1-m high with a field of about 0.8~T.
They are assumed empty in this study.
The spoilers are extended into the tunnel wall/dirt by 60 cm horizontally to prevent muon
backscattering. The spoiler geometry, magnetic field distribution as well as simulated particle
tracks in the spoiler region -- as modeled with MARS15 -- are shown in Fig.~2.
The calculations are performed for two cases: with and without muon spoilers in the BDS tunnel.

SiD detector response is calculated using the Simulator for the Linear Collider (SLIC)
and its geometry package, Linear Collider Detector Description (LCDD). LCDD includes various 
detector configurations, such as SiD, GLD, TESLA(D09) and others~\cite{slic}. Simulations in this
paper are performed using the Silicon Detector (SiD) geometry. SLIC
takes into account a detailed description of the SiD geometry, the magnetic field and sensitivity
of different sub-systems. A two-dimensional view of a SiD quadrant
is shown in Fig.~3. SLIC provides possibility to calculate 
time and space distributions of hits in the detector. The SiD LCDD description includes 12 detector sub-systems.
``Calorimeter'' type hits are simulated in the Muon Endcap and Barrel, Hcal Endcap
and Barrel, Ecal Endcap and Barrel, ForwardEcal Endcap and Luminosity Monitor. ``Tracker'' hits are simulated for tracker
(Endcap and Barrel) and vertex (Endcap and Barrel) detectors. 
The ``calorimeter'' hit information is quantized into cells in the GEANT4 program. 
The total energy deposited (and time of deposition) by each primary particle
in a calorimeter cell is recorded. This, however, is done at a finer segmentation than is 
expected to be available in a real detector. A ``real'' number of hits could be lower if
realistic segmentation and thresholds are taken into account. It could be done
when details of the detector geometry are finalized.

The nominal ILC beam parameters~\cite{ilc-par} are used in this study: the beam consists of 5 trains
per second with 2820 bunches in each train, time between trains is 199 ms, the train length is 868 $\mu s$. There are 
$2 \times 10^{10}$ positrons/electrons in each bunch and the luminosity is $2 \times 10^{34}$ cm$^{-2}$ s$^{-1}$.

\section{BDS Induced Detector Backgrounds}

The spoilers drastically reduce muon and electron fluxes at the detector. The BDS-induced muon
flux averaged over the tunnel cross-section at the entrance to the experimental hall is
4.1~cm$^{-2}s^{-1}$ without spoilers, while it is 1.2$\times$10$^{-3}$~cm$^{-2}s^{-1}$ with
the spoilers described in the previous section. Note that the effect of muon penetration
through the central gap and holes in the left/right parts is quite substantial: filling them
with steel reduces the above flux by a factor of three, resulting in a shielding effect
of such hypothetical spoilers of a factor of 10$^4$. Filling/screening of these openings (at least partial)
to reduce further backgrounds in the detector and radiation levels the experimental halls
should be considered in the future.

Average number of background particles produced by the positron beam and their average energy
at the Muon Endcap (589 cm from the IP) are presented in Tables 1-2.
Energy spectra of background particles are shown in Fig.~4.
Average number and energy of photons 
and positrons are practically not changed by the spoilers. Most of photons and positrons
are coming near or inside the beam pipe while radial distributions of other particles
are rather flat over the first three meters from the beam axis (see Fig.~5). 
Based on a limited statistics for neutrons, their flux rises about 20 times
with the spoilers installed. The neutrons coming from the tunnel are not a serious concern at this
stage, because it is envisioned that there will be standard concrete wall plugging
the tunnel at the entrance to the experimental hall which will absorb most of the BDS neutrons.
Similar machine-induced backgrounds irradiate the other side of the detector from the electron beam.

There is also the IP-related background in the detector, $e^+ e^-$ pairs and radiative Bhabhas 
from beam-beam interactions~\cite{snow05}. Maruyama~\cite{maruyama} has calculated responses of
the vertex and tracker SiD sub-detectors  to these backgrounds using the GEANT3 and Guineapig programs.
These backgrounds depend on the beam crossing angle.  The IP-related background could be reduced 
using a low-Z masks in the detector. The 20-mrad option with a low-Z shielding is selected for
comparison with calculations presented in this report. The BDS-induced background hit rates are
compared with hits produced by electron-positron interactions in the IP.  
Secondary particles from 250$\times$250~GeV $e^+ e^-$ collisions are simulated using 
the PYTHIA code~\cite{pythia} with a cross-section of $1.8 \times 10^{-8}$~mb. 
The detector response for these particles is calculated using the SLIC code. 

\section{Hit Rates in Sub-Detectors}

Hit rates in SiD detector sub-systems from the positron tunnel, IP backgrounds and  $e^+ e^-$ events
are presented in Fig.~6. Background from the tunnel (no spoiler option) produces
much more hits in the muon system than $e^+ e^-$ interactions in the IP. The BDS-backgrounds
(without spoilers) and $e^+ e^-$ collisions give almost identical hit rates in the hadronic calorimeter.
The spoilers reduce the rates in these sub-detectors by more than three orders of magnitude.
In all other sub-detectors, the $e^+ e^-$ contribution dominates.

The muon spoilers reduce the BDS-induced backgrounds in most of the sub-systems. 
The only exception is the vertex detectors where the effect is opposite, although
this conclusion is based on a low statistics for these sub-detectors.
The effect of spoilers should be negligible there, because
the main source of the tunnel background for vertex detectors is near-beam positrons and photons.
These positrons are due to the beam halo ``quasi-elastic'' scattering in the collimator jaws. They pass
the spoilers within the beam pipe. The related photons are created after the spoilers. 
Therefore, the numbers and spectra of these particles in the near-beam region at the IP
(vertex detectors) are practically not affected by the tunnel spoilers
(see Tables~1-2 and Fig.~4). As seen from Fig.~6, the IP-induced backgrounds dominate 
the vertex Endcap and Barrel occupancies.

A statistical uncertainty of neutron-produced hit rates is substantial. At this stage,
the details that drive the neutron fluxes in the ILC detectors -- configuration, dimensions and materials
of the experimental hall and tunnel interface, passive materials of the calorimeters -- are quite uncertain.
There are also concerns about accuracy of the current SLIC/GEANT4 simulation of low-energy neutron
transport \cite{lemrani}.

Hit rates for the BDS backgrounds are presented in Tables~3-4. Muons are the main source
of the machine backgrounds for the SiD except luminosity monitor.
Most of the muons have enough energy to pass through the whole detector (see Fig.~7). They move perpendicularly
to the sensitive layers of the Endcaps (Muon, Hcal, Ecal, FEcal).
Therefore, every muon produces about one hit in a sensitive layer of the Endcaps. Total hit number $N_{hit}$
in the Endcaps can be estimated as
\begin{equation}
N_{hit} = \int_{r_{min}}^{r_{max}} dr f_{\mu}(r) \cdot N_{sl},
\end{equation}
where $f_{\mu}(r)$ is the radial distribution of incoming muons (Fig.~4), $r_{min}$ and $r_{max}$ are
the minimal and maximal radii of the Endcap, and $N_{sl}$ is a number of sensitive layers. 
The tracker Endcap is the special case. It consists of two sub-systems, 5 layers each, and
 every layer has different minimal and maximal radii. There are two detection planes in each layer. In this case
\begin{equation}
N_{hit} = 2 \times 2 \times \sum_{n=1}^{5} \int_{r_{min(n)}}^{r_{max(n)}} dr f_{\mu}(r) .
\end{equation}

A comparison of simulation and a simple model (1)-(2) is shown in Table~5.
The model agrees with the SLIC simulations of muon hits in the Endcaps within about 20\%. 
This model can be used to estimate hit rates in the Endcaps for different Endcap designs.

\section{Tolerable Limits and Machine Backgrounds}

Possible approach to the tolerable background levels in different ILC detector sub-systems was
discussed in Ref.~\cite{kozanecki}. For calorimeter, tracker and vertex detectors, a limit
on a background occupancy was estimated to be about 1\%. 
A segmentation of the SiD calorimeters is not finalized yet. Using a cell size of 1 cm$^2$
and Table 3, the background occupancy for the calorimeters could be estimated. Results are presented in Table~6
and can be simply re-scaled to another cell size. The occupancy levels are smaller than 1\% if
the detector integration time is shorter than a time between bunches. 

An estimate of acceptable background levels in the SiD tracker has also been presented in Ref.~\cite{kozanecki2}. 
To avoid a pattern recognition problem, the hit density from charged particles should be lower than 
0.2 hit/cm$^2$/bunch. To avoid a pile-up problem, the background level should be lower than 0.2 hit/mm$^2$/train. 
The calculated tunnel-related background distributions in the Tracker Endcap and Barrel are rather flat (see Fig.~8).
Therefore, 
the hit densities are simply the ratios of hit numbers  (from Table~3) and areas of the Endcap and Barrel sensitive layers,
respectively.
For the Tracker Endcap, the hit density is $7 \times 10^{-4}$/cm$^2$/bunch or 0.02/mm$^2$/train. For the Tracker Barrel, the 
hit density is  $4 \times 10^{-5}$/cm$^2$/bunch or 0.001/mm$^2$/train. The tunnel background (no spoiler case) 
in the SiD tracker is also lower than the acceptable levels as defined in Ref.~\cite{kozanecki2}.

There are two estimates of the acceptable background levels in the SiD muon system~\cite{maruyama2}.
The RPCs (sensitive media) need 1~ms
to re-charge a 1 cm$^2$ area around the avalanche. Therefore, the 
 background rates of the order of 100 Hz/cm$^2$  would result in an unmanageable dead
time. A radial hit distribution in the Muon Endcap is shown in Fig.~9. There are 14100 bunches/s,
thus the tunnel background rate in the Muon Endcap  (without spoilers)
is about 400~Hz/cm$^2$, four times larger than the acceptable level~\cite{maruyama2}.
The other limit (1~muon/cm$^2$/s) was presented as a conservative expert estimate~\cite{maruyama2}.
A radial muon distribution at the Muon Endcap entrance is presented in Fig.~10. 
The tunnel backgrounds (without spoilers) exceed this level about four times.

If the detector sensitivity window is less than the time between the bunches, it is possible to use 
the difference in the signal and background timing to increase the signal/background ratio. 
Time distributions of hits in the detector sub-systems are presented in Fig.~11-34. 
The time of a bunch crossing is chosen to be zero for these plots. A sub-detector starts collecting signals
after a bunch crossing.
Background hits produced before crossing do not count. The hits created by the tunnel background after a bunch
crossing  are presented in Table~7 together with the total rates.
The machine background after a bunch crossing is about twice lower than the total for the subsystems
where muons dominate. Time window could be a very effective suppressor of background for the barrels
(Muon Barrel, Hcal Barrel, Ecal Barrel). Note again that Figs.~1, 3-4, 6-8, 11-34 and Tables 1-5 present 
the machine background coming from the positron side only. About the same number of background particles comes
from the electron side. Estimates of the occupancy (Table 6), comparison of background levels and tolerable limits
in the muon system and tracker are performed for particles coming from the both sides.

The machine-related backgrounds are calculated in this study for the positron beam coming
to the IP. The muon fluxes here are slightly higher than for the electron-beam side,
because of an extra annihilation $e^+ e^- \rightarrow \mu^+\mu^-$ at the very beginning
of shower development in the collimators. The choice of collimator
materials is important. A cross-section of the above annihilation process is 
proportional to the atomic charge $Z$. A cross section of a muon pair production
in $\gamma A$ interactions (Bethe-Heitler processes -- the dominant source of muon fluxes
at the detector) rises as $Z^2$. Therefore, the BDS-generated muon fluxes in the collision
halls can be reduced by use of low-$Z$ material for the collimators contributing most
to the background. In this case, the difference between the positron and electron side backgrounds
will be more significant because of a more visible contribution from annihilation
in low-$Z$ material.

\section{Backgrounds and Detector Performance}

Backgrounds affect ILC detector performance in three major ways: detector radiation aging,
reconstruction of background objects (for example, tracks) not related to products of $e^+ e^-$
interactions, and deterioration of detector resolution (for example, jets energy resolution due to extra
energy from background hits). Detailed simulations (beyond calculations presented in this paper) of 
the detector response to particles from primary $e^+ e^-$ collisions as well as other sources of 
backgrounds are needed in order to select final configuration of the BDS, including shielding, 
and to optimize detector performance.

Analysis of fluxes presented on Fig.~6 demonstrates that tunnel backgrounds provide large number
of extra hits in the muon detectors, while in other detectors backgrounds from the IP dominate.
Still, even without magnetic spoilers, background muon fluxes are within tolerable levels for 
muon detectors designed for modern collider experiments~\cite{d0-2005}. With magnetic iron spoilers,
the BDS-related backgrounds in all SiD detectors become well below the IP and e+e- interaction
backgrounds improving detector longevity and performance. Note that another essential
function of the tunnel spoilers is to reduce radiation levels in a second experimental hall
where construction work on a second collider detector can be underway while the
ILC beams are on.

\section{Conclusions}

Detailed calculations of the background fluxes in the SiD detector components for ILC parameters
from [11] and assumption of 0.1\% beam loss in BDS are presented in this paper.
Background flux distributions \textit{vs} distance to the beam pipe, type of the particle
creating energy deposition and timing of the hits with respect to the bunch crossing are presented.
These studies provide important information for ILC detector designers opening options to reduce
backgrounds by appropriate selection of detector properties, such as sensitivity to different
types of particles and timing characteristics of the detectors. Option of reducing muon fluxes
on the ILC detector by installing magnetic iron spoilers in the BDS tunnel is discussed.
Such spoilers will reduce background muon fluxes on the ILC detector components to the level
well below backgrounds from the IP region and $e^+ e^-$ collisions.

\section{Acknowledgements}

The authors are thankful to Marcel Demarteau and Alexander Drozhdin for fruitful discussions.

\newpage

\begin{description}
\item[Table. 1]: {Average number of particles per bunch at the SiD from positron tunnel.}
\end{description}

\begin{center}

\begin{tabular}{llllll}
\hline
               &$\gamma$  &$\mu^\pm$  & $e^+$  & $e^-$               &   neutron   \\
\hline
With spoilers  & 2927     &0.024     & 1172   & $3.6 \cdot 10^{-4}$ &6364         \\
No spoilers    & 2942     &60.4      & 1095   & 10                  &346          \\
\hline
\end{tabular}
\end{center}

\begin{description}
\item[Table. 2]: {Average kinetic energy (GeV) of particles  at the SiD from positron tunnel.}
\end{description}
\begin{center}
\begin{tabular}{llllll}
\hline
  & $\gamma$           & $\mu^\pm$ &  $e^+$  & $e^-$               &   neutron   \\
\hline
with spoilers & $5.4 \cdot 10^{-3}$ & 38        & 251    & 0.13  &   $1.6 \cdot 10^{-3}$\\
no spoilers   & $5.5 \cdot 10^{-3}$ & 28       & 250    & 0.19  &   $7   \cdot 10^{-4}$\\
\hline
\end{tabular}
\end{center}

\begin{description}
\item[Table. 3]: {Contribution of particles from the positron tunnel to hit rates in the SiD sub-detectors without spoilers.}
\end{description}

\begin{center}

\begin{tabular}{llllllll}
\hline
              & All                 & $\gamma$ & $\mu^\pm$ & $e^+$  & $e^-$& neutron  \\
              & hits/bunch          & \%       & \%        & \%     & \%   & \%       \\   
\hline
Muon Endcap   & 4711                & 0.2      & 99.4      & 0.1    & 0.3  & 0      \\
Muon Barrel   & 49                  & 0        & 100       & 0      & 0    & 0    \\
Hcal Endcap   & 584                 & 0        & 100       & 0      & 0    & 0    \\
Hcal Barrel   & 314                 & 0        & 100       & 0      & 0    & 0    \\
Ecal Endcap   & 435                 & 0        & 100       & 0      & 0    & 0    \\
Ecal Barrel   & 100                 & 0        & 100       & 0      & 0    & 0    \\
FEcal Endcap  & 12                  & 0        & 100       & 0      & 0    & 0    \\
Tkr Endcap    & 79                  & 0        & 95        & 5      & 0    & 0    \\
Tkr Barrel    & 20                  & 59       & 41        & 0      & 0    & 0   \\
Vtx Endcap    & $6.7 \cdot 10^{-3}$ & 0        & 100       & 0      & 0    & 0  \\
Vtx Barrel    & $5.4 \cdot 10^{-3}$ & 0        & 100       & 0      & 0    & 0  \\
Luminosity Monitor & 36             & 45       & 10        & 45     & 0    & 0     \\
\hline
\end{tabular}
\end{center}

\begin{description}
\item[Table. 4]: {Contribution of particles from the positron tunnel to hit rates in the SiD sub-detectors with spoilers.}
\end{description}

\begin{center}

\begin{tabular}{llllllll}
\hline
              & All          & $\gamma$  & $\mu^\pm$ & $e^+$ & $e^-$ & neutron  \\
              & hits/bunch   & \%       & \%       & \%      & \%  &  \%       \\   
\hline
Muon Endcap   & 2.4         & 0        & 99        & 0        & 0  & 1   \\
Muon Barrel   & 0.045       & 0        & 100       & 0        & 0  & 0      \\
Hcal Endcap   & 0.642       & 0        & 100       & 0        & 0  & 0      \\
Hcal Barrel   & 0.074       & 0        & 100       & 0        & 0  & 0    \\
Ecal Endcap   & 1.28        & 62       & 38        & 0        & 0  & 0       \\
Ecal Barrel   & 0.41        & 98.4     & 1.6       & 0        & 0  & 0       \\
FEcal Endcap  & $5.4 \cdot 10^{-4}$    & 0         & 100      & 0  & 0      & 0  \\
Tkr Endcap    & 10.5        & 72.5        & 0.7       & 26.8     & 0  & 0        \\
Tkr Barrel    & 4           & 70        & 0         & 30       & 0  & 0       \\
Vtx Endcap    & 1.6         & 100        & 0         & 0        & 0  & 0  \\
Vtx Barrel    & 0.8         & 0        & 0         & 100      & 0  & 0  \\
Luminosity Monitor & 36     & 35       & 0         & 65       & 0  & 0    \\
\hline
\end{tabular}
\end{center}

\newpage
\begin{description}
\item[Table. 5]: {Comparison between simple model (1)-(2) and simulation.}
\end{description}

\begin{center}

\begin{tabular}{lllllll}
\hline
              &spoilers &  $N$ muons             & $N$ layers      & muon hits& muon hits  & all hits  \\
              &         & $r^{min}<r<r^{max}$    &                 & (1)-(2)  & simulation & simulation \\   
\hline
Muon Endcap   &  no     & 60.4     & $48 \cdot 2$ & 5798   & $4685 \pm 160$ & 4711      \\
Hcal Endcap   &  no     & 10.7     & $34 \cdot 2$ & 725    & $584  \pm 50$  & 584       \\
Ecal Endcap   &  no     & 9.5      & $30 \cdot 2$ & 568    & $435  \pm 43$  & 435       \\ 
FEcal Endcap  &  no     & 0.078    & $30 \cdot 2$ & 4.7    & $11.7 \pm 4.6$ & 11.7       \\ 
Tkr Endcap    &  no     & 23       & $(5)2 \cdot 2$ & 92     & $75   \pm 10$  & 79       \\ 
Lum Monitor   &  no     & 0.024    & $50 \cdot 2$ & 2.4    & $3.9  \pm 2.5$ & 36       \\ 
Muon Endcap   &  yes    & 0.024    & $48 \cdot 2$ & 2.3   & $2.4  \pm 0.6$ & 2.4      \\
Hcal Endcap   &  yes    & $6.8\cdot 10^{-3}$ & $34 \cdot 2$& 0.46    & $0.64  \pm 0.28$  & 0.64       \\
Ecal Endcap   &  yes    & $6.8\cdot 10^{-3}$ & $30 \cdot 2$& 0.41    & $0.48  \pm 0.22$  & 1.28       \\ 
FEcal Endcap  &  yes    & $2.2\cdot 10^{-5}$ & $30 \cdot 2$&  $1.3\cdot 10^{-3}$ & $5.4 \cdot 10^{-4}  \pm 4.4 \cdot 10^{-4}$  & $5.4 \cdot 10^{-4}$ \\ 
Tkr Endcap    &  yes    & 0.013              & $2 \cdot 2$ & 0.052   & $0.078   \pm 0.040$  & 10.5       \\ 
Lum Monitor   &  yes    & $1.3\cdot 10^{-5}$ & $50 \cdot 2$&  $1.3\cdot 10^{-3}$ & $3.4 \cdot 10^{-4}  \pm 2.1 \cdot 10^{-4}$  & 36. \\

\hline
\end{tabular}
\end{center}

\begin{description}
\item[Table. 6]: {Tunnel background occupancies in sub-detectors (no spoilers) taking
into account both electron and positron beam losses.}
\end{description}

\begin{center}

\begin{tabular}{llll}
\hline
              & Sensitive area   & Hit number      & occupancy   \\
              & cm$^2$           & per bunch       & per bunch   \\   
\hline
Muon Endcap   & 1.3 $\cdot 10^8 $ & 4711 $\cdot 2$ & 0.008 \%     \\
Muon Barrel   & 8.2 $\cdot 10^7 $ & 49   $\cdot 2$ & 0.0001\%     \\
Hcal Endcap   & 3.9 $\cdot 10^6 $ & 584  $\cdot 2$ & 0.03 \%      \\
Hcal Barrel   & 2.2 $\cdot 10^7 $ & 314  $\cdot 2$ & 0.003 \%     \\
Ecal Endcap   & 2.9 $\cdot 10^6 $ & 435  $\cdot 2$ & 0.03 \%      \\
Ecal Barrel   & 9.0 $\cdot 10^6 $ & 100  $\cdot 2$ & 0.002 \%     \\
FEcal Endcap  & 1.0 $\cdot 10^5 $ & 12   $\cdot 2$ & 0.02 \%      \\
Lum Monitor   & 6.3 $\cdot 10^4 $ & 36   $\cdot 2$ & 0.12 \%      \\
\hline
\end{tabular}
\end{center}

\begin{description}
\item[Table. 7]: {Tunnel background in SiD sub-detectors, total and after bunch crossing (BC)
taking into account both electron and positron sides.}
\end{description}

\begin{center}

\begin{tabular}{lllll}
\hline
              & no spoilers         & no spoilers          & with spoilers        & with spoilers \\ 
\hline
              & total               & after BC             & total                & after BC     \\
              & hits/bunch          & hits/bunch           & hits/bunch           & hits/bunch   \\
\hline
Muon Endcap   & 9422                &  3646                & 4.76     & 2.7  \\
Muon Barrel   & 98                  &  48                  & 0.045    & 0.018 \\ 
Hcal Endcap   & 1168                &  512                 & 0.642    & 0.341 \\ 
Hcal Barrel   & 628                 &  322                 & 0.148    & 0.060 \\   
Ecal Endcap   & 870                 &  404                 & 2.56     & 2.046 \\   
Ecal Barrel   & 200                 &  102                 & 0.82     & 0.806 \\ 
FEcal Endcap  & 24                  &  13.4                & $1.1\cdot 10^{-4}$ & $5.9\cdot 10^{-4}$ \\
Tkr Endcap    & 158                 &  84                  & 21       & 16.89\\ 
Tkr Barrel    & 40                  &  34                  & 8        & 8 \\
Vtx Endcap    & $1.3 \cdot 10^{-2}$ &  $1.3 \cdot 10^{-2}$ & 3.2      &3.2\\ 
Vtx Barrel    & $1.1 \cdot 10^{-2}$ &  $1.1 \cdot 10^{-2}$ & 1.6      &1.6\\ 
Luminosity Monitor & 72             &  39.4                & 72       &20 \\ 
\hline
\end{tabular}
\end{center}

\newpage
\begin{figure}[hbt!]
\centering\epsfig{figure=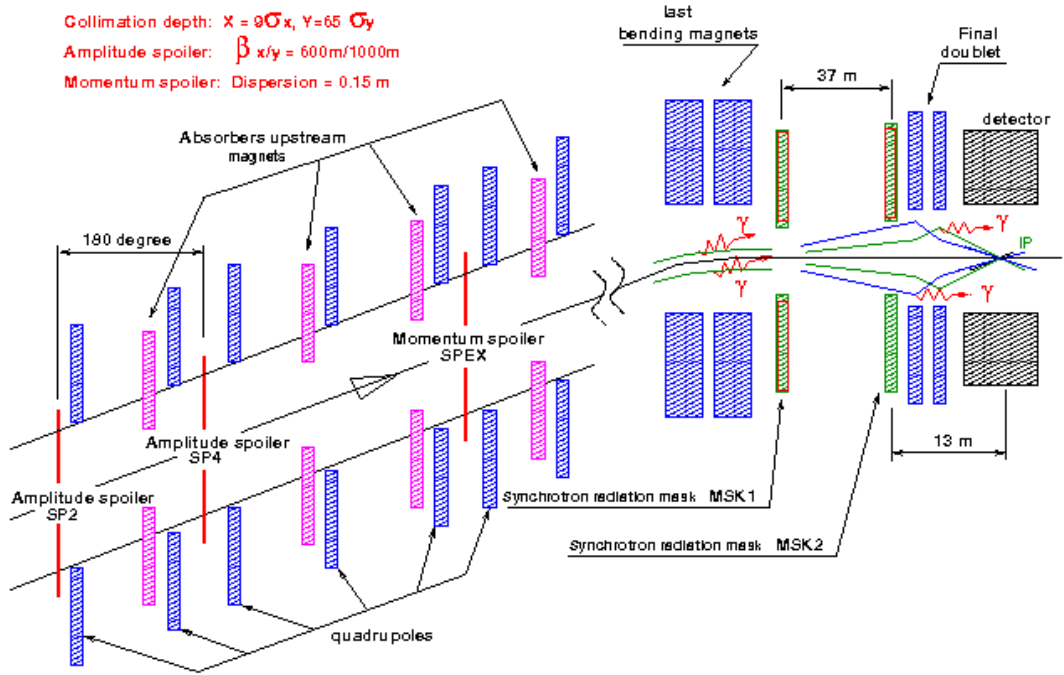,width=1.0\linewidth}
\centering\epsfig{figure=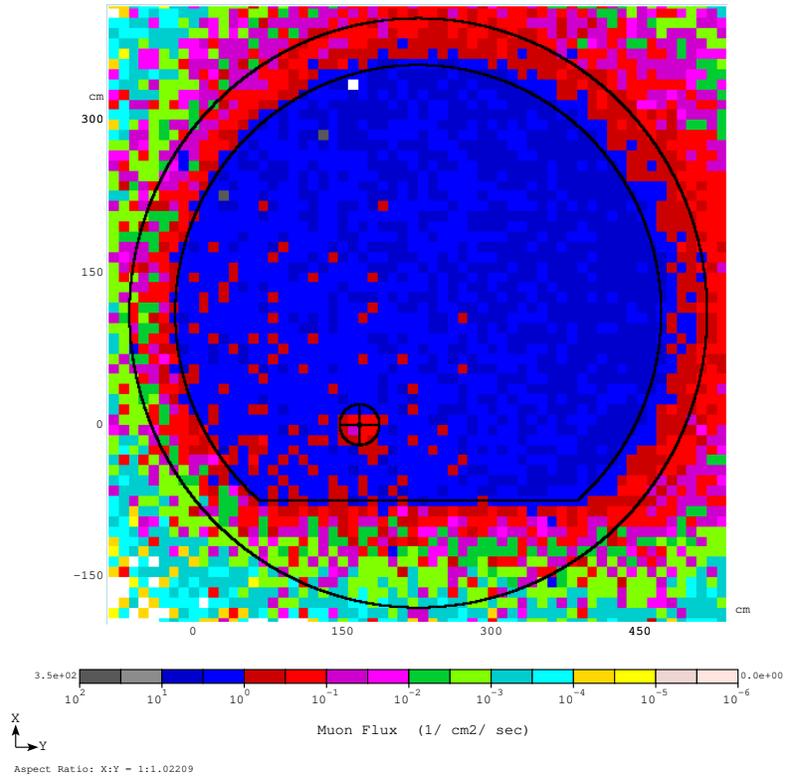,width=0.7\linewidth}
\caption{BDS layout (top) and muon flux at the tunnel-experimental hall transition (bottom).}
\end{figure}

\newpage
\begin{figure}[hbt!]
\centering\epsfig{figure=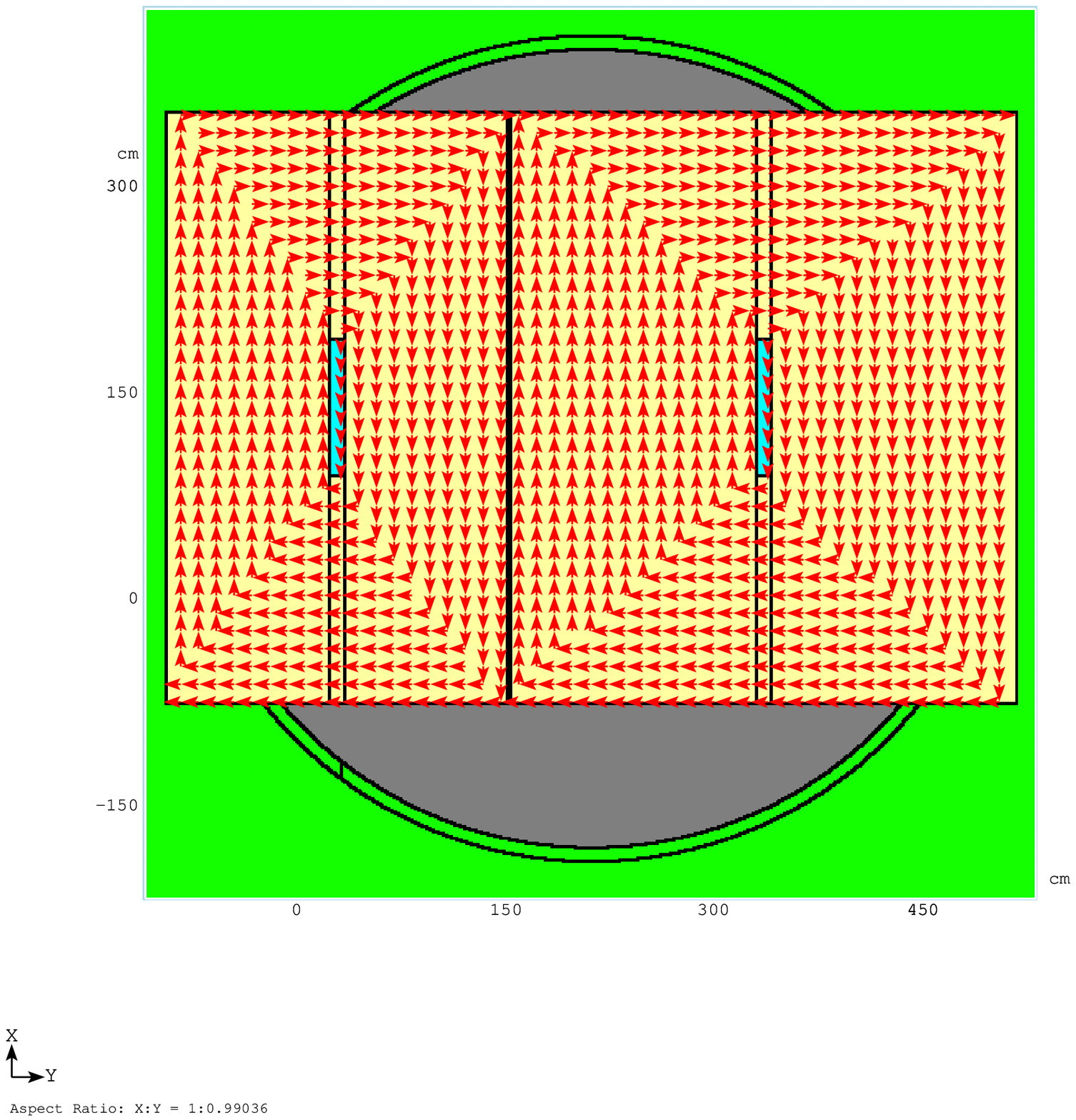,width=0.67\linewidth}
\centering\epsfig{figure=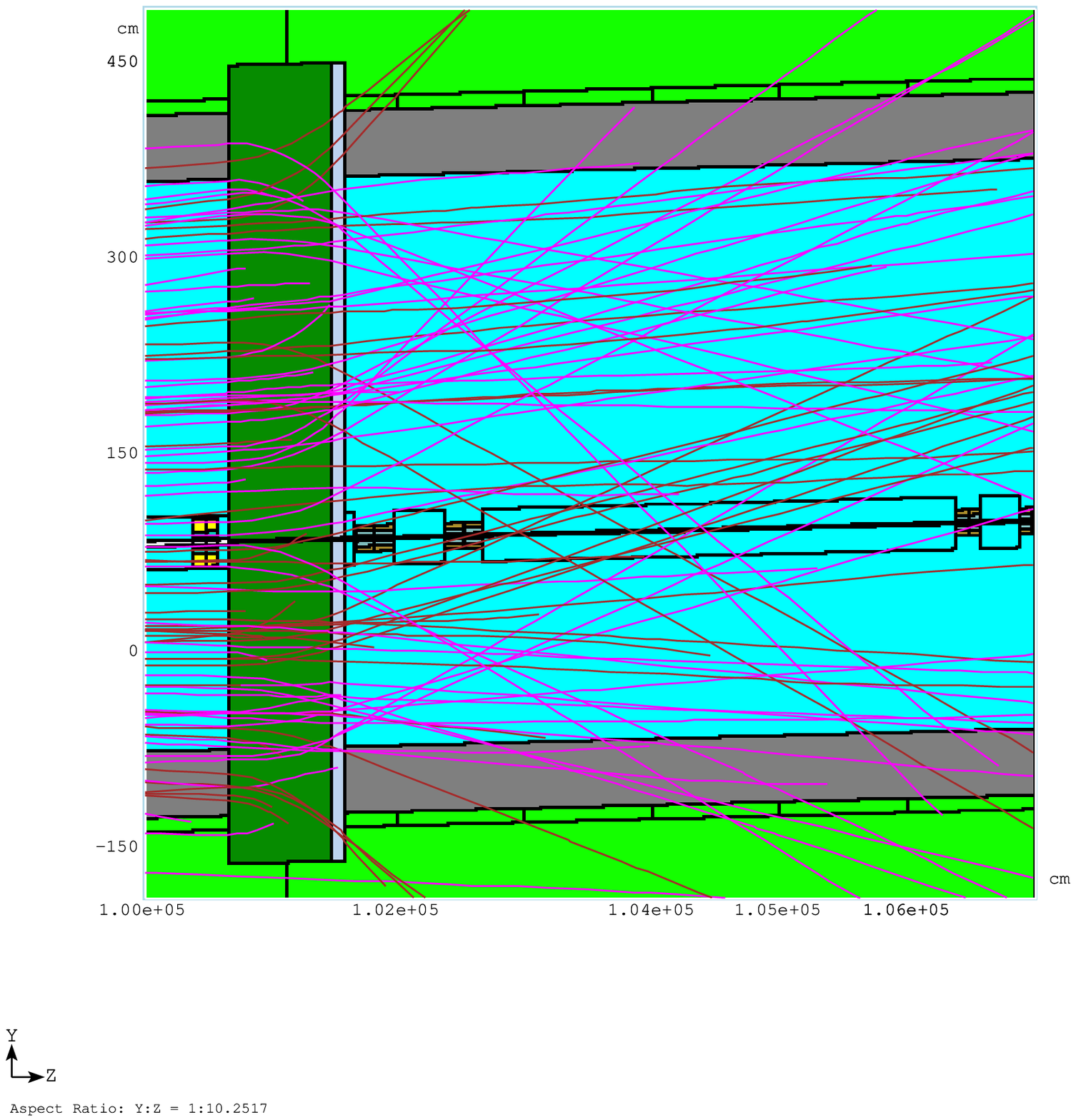,width=0.67\linewidth}
\caption{MARS15 muon spoiler with magnetic field lines (top) and muon tracks in the spoiler region (bottom).}
\end{figure}

\newpage
\begin{figure}[hbt!]
\vspace{-2.0cm} 
\centering\epsfig{figure=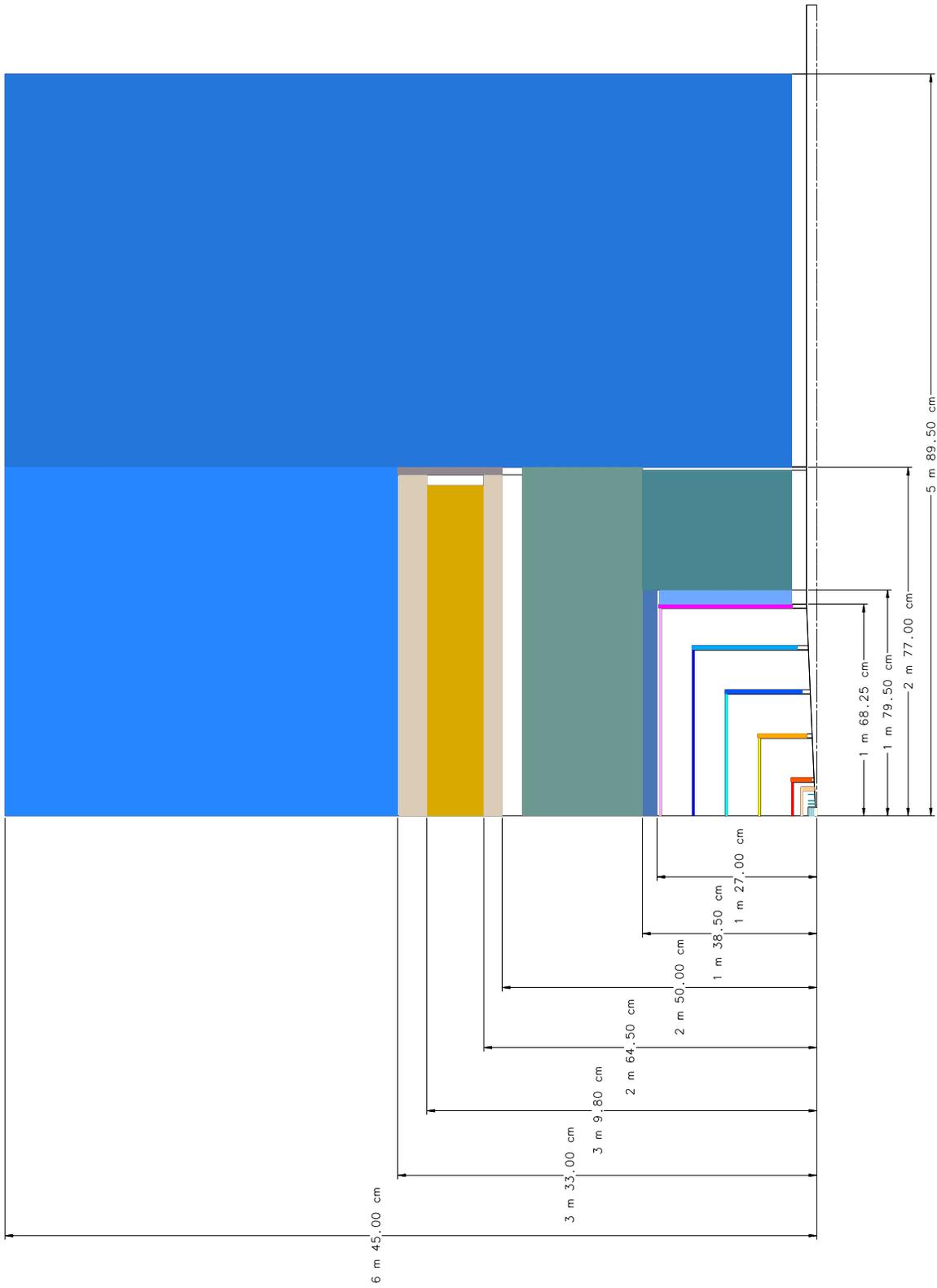,width=1.0\linewidth}
\caption{Quarter section of Silicon Detector (SiD).}
\end{figure}

\newpage
\begin{figure}[hbt!]
\centering\epsfig{figure=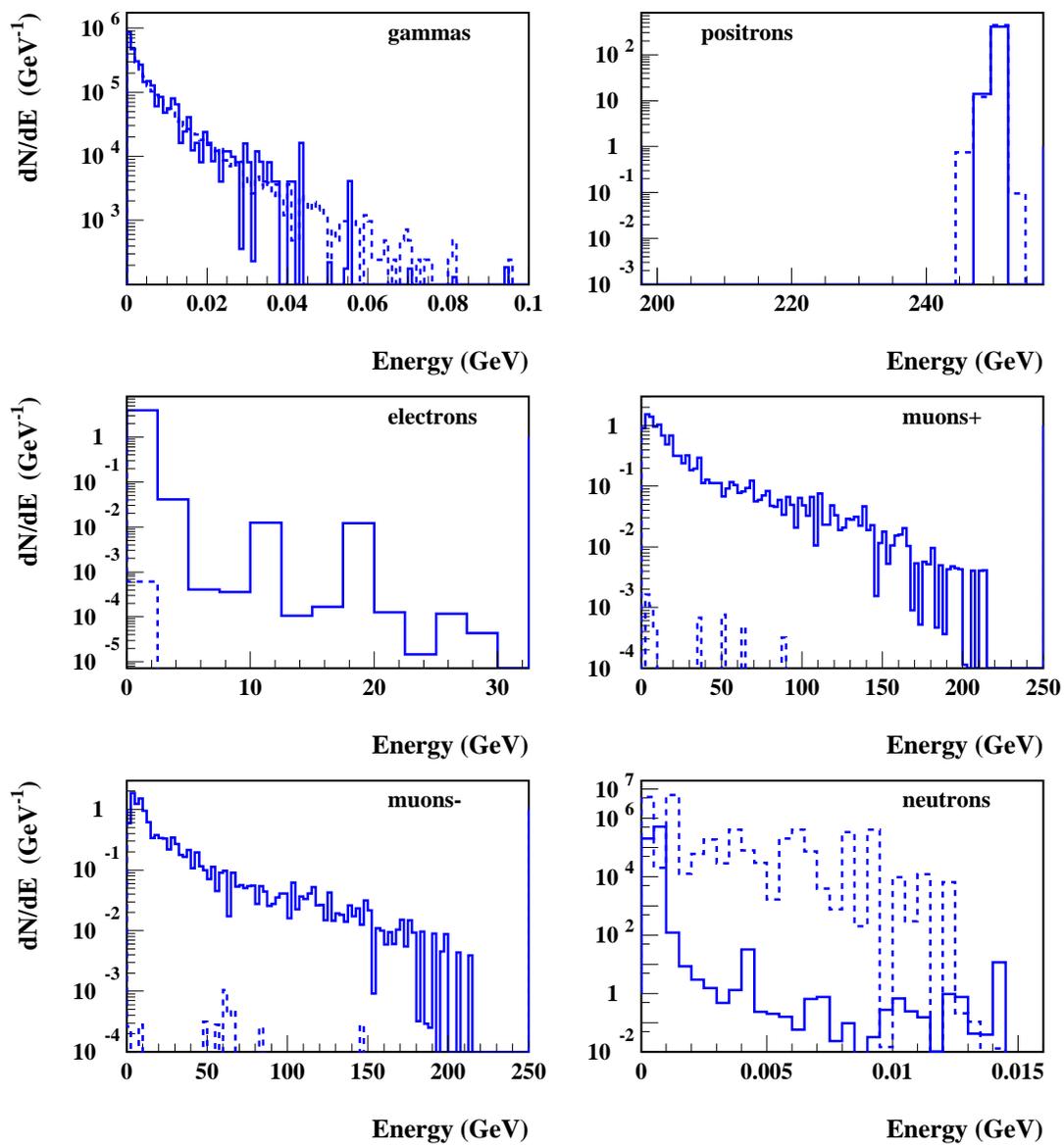,width=1.0\linewidth}
\caption{Energy spectra of particles at the SiD detector (per bunch). Solid line - no spoilers, 
dashed line - tunnel with spoilers. Particles come from positron tunnel only.}
\end{figure}

\newpage
\begin{figure}[hbt!]
\centering\epsfig{figure=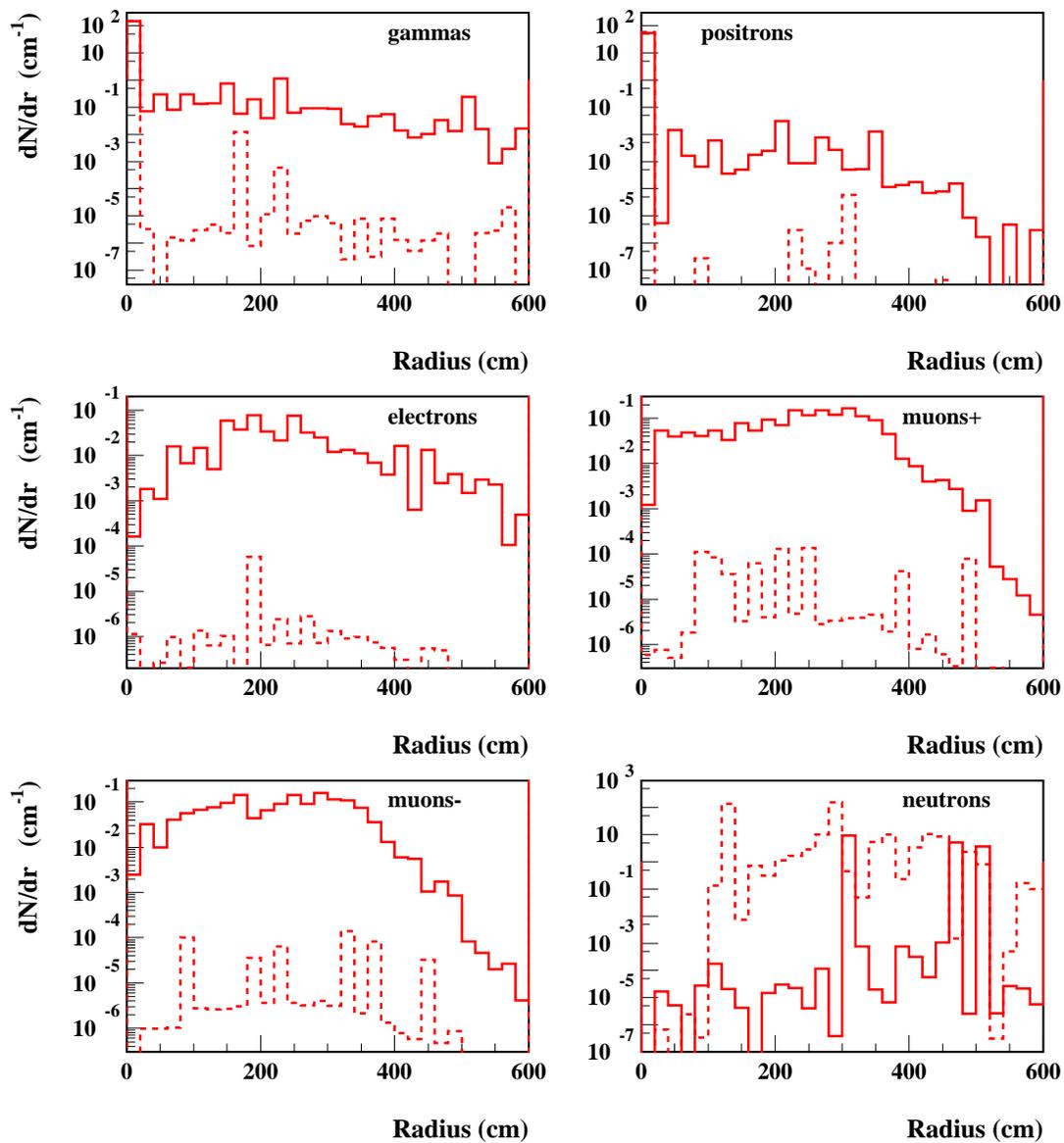,width=1.0\linewidth}
\caption{Radial distributions of particles at the SiD detector. 
Solid line - no spoilers, dashed line - tunnel with spoilers. Particles come from positron tunnel only. }
\end{figure}

\newpage
\begin{figure}[hbt!]
\centering\epsfig{figure=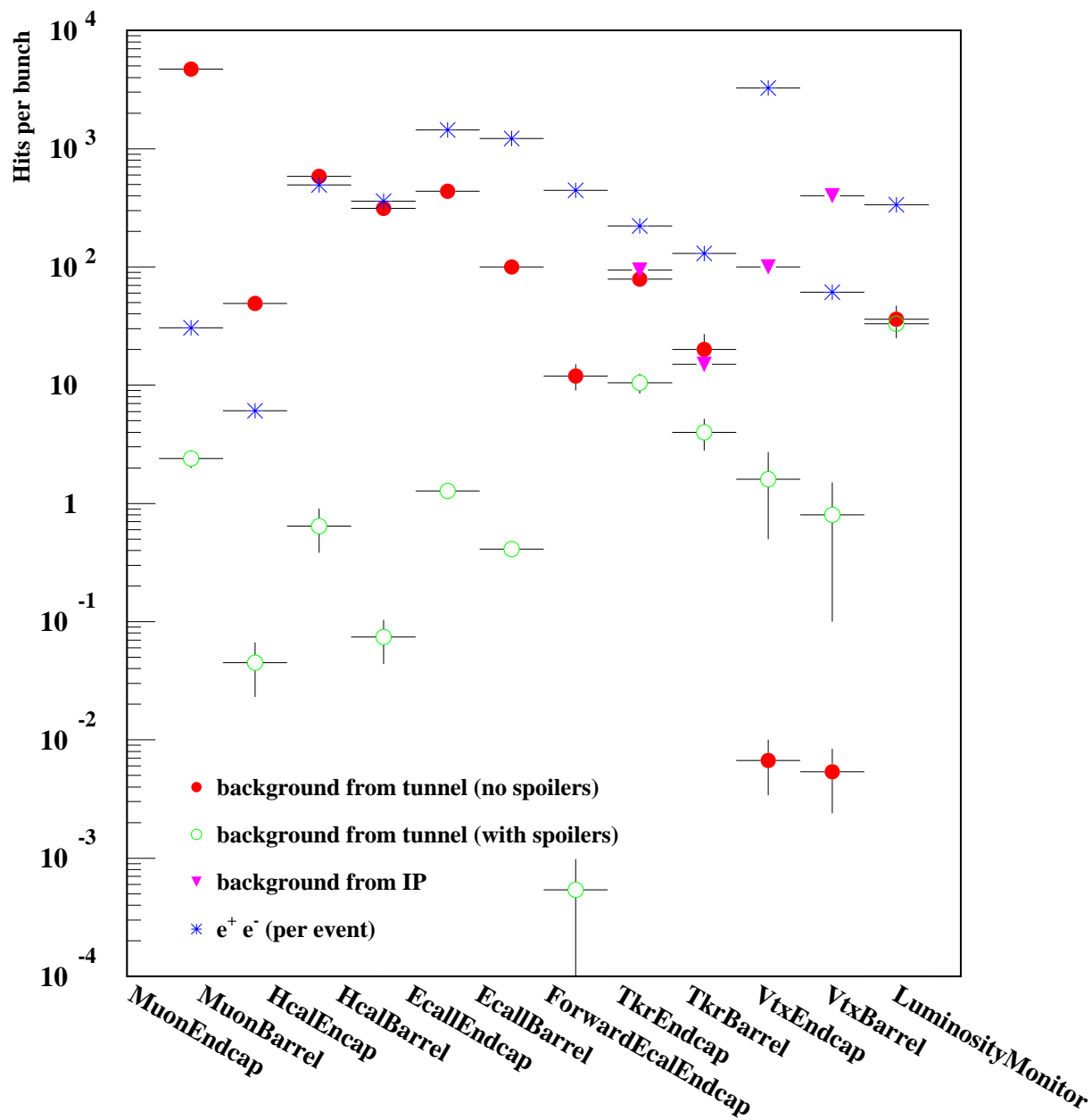,width=1.0\linewidth}
\caption{Hit rates in different detector subsystem. Tunnel background is created by particles coming from positron tunnel only. } 
\end{figure}

\newpage
\begin{figure}[hbt!]
\begin{minipage}[b]{0.95\linewidth}
\centering\epsfig{figure=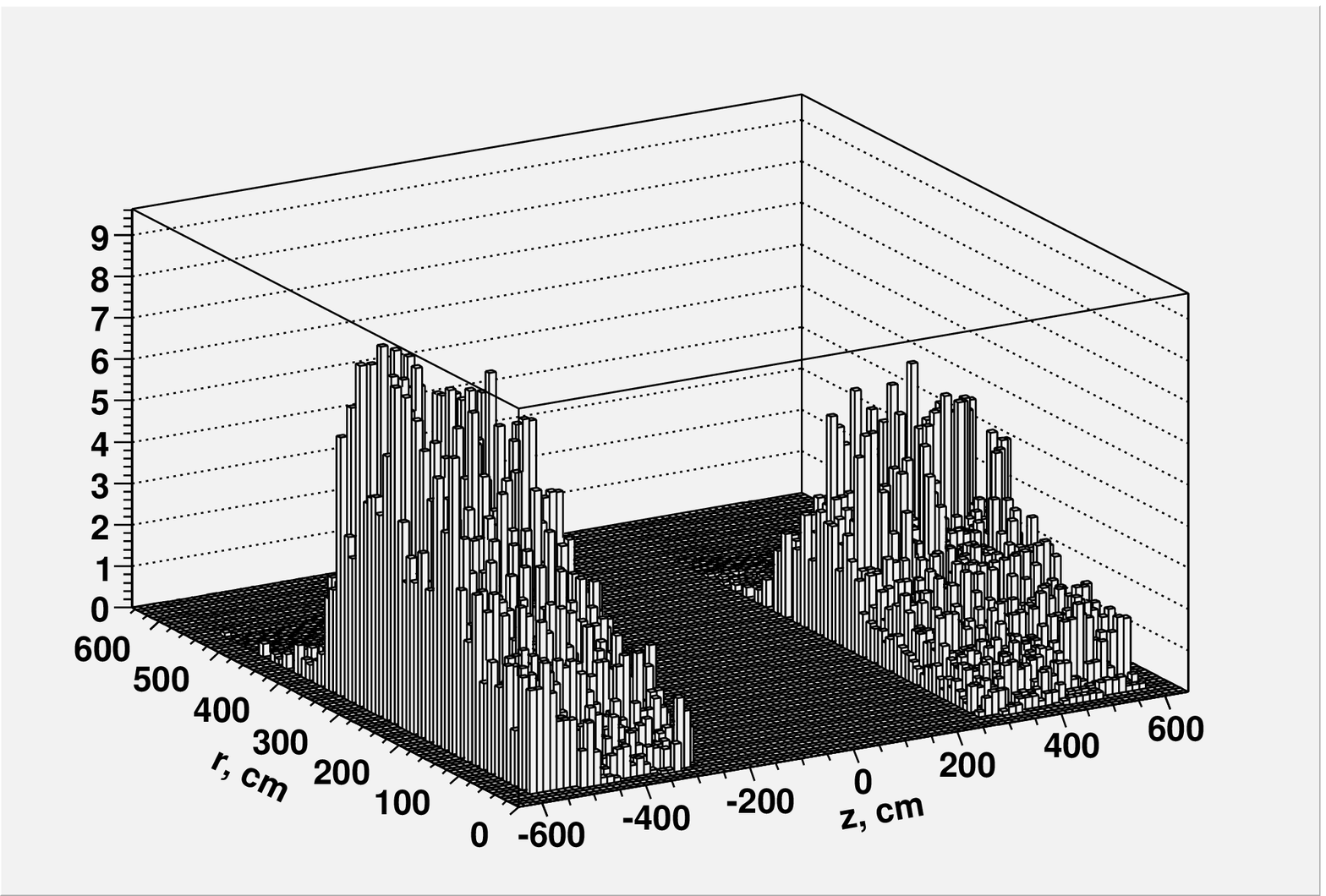,width=\linewidth}
\caption{RZ distribution of hits per bunch in Muon Endcap. No spoilers. Background is created by particles coming from positron tunnel only. }
\end{minipage}

\begin{minipage}[b]{0.95\linewidth}
\centering\epsfig{figure=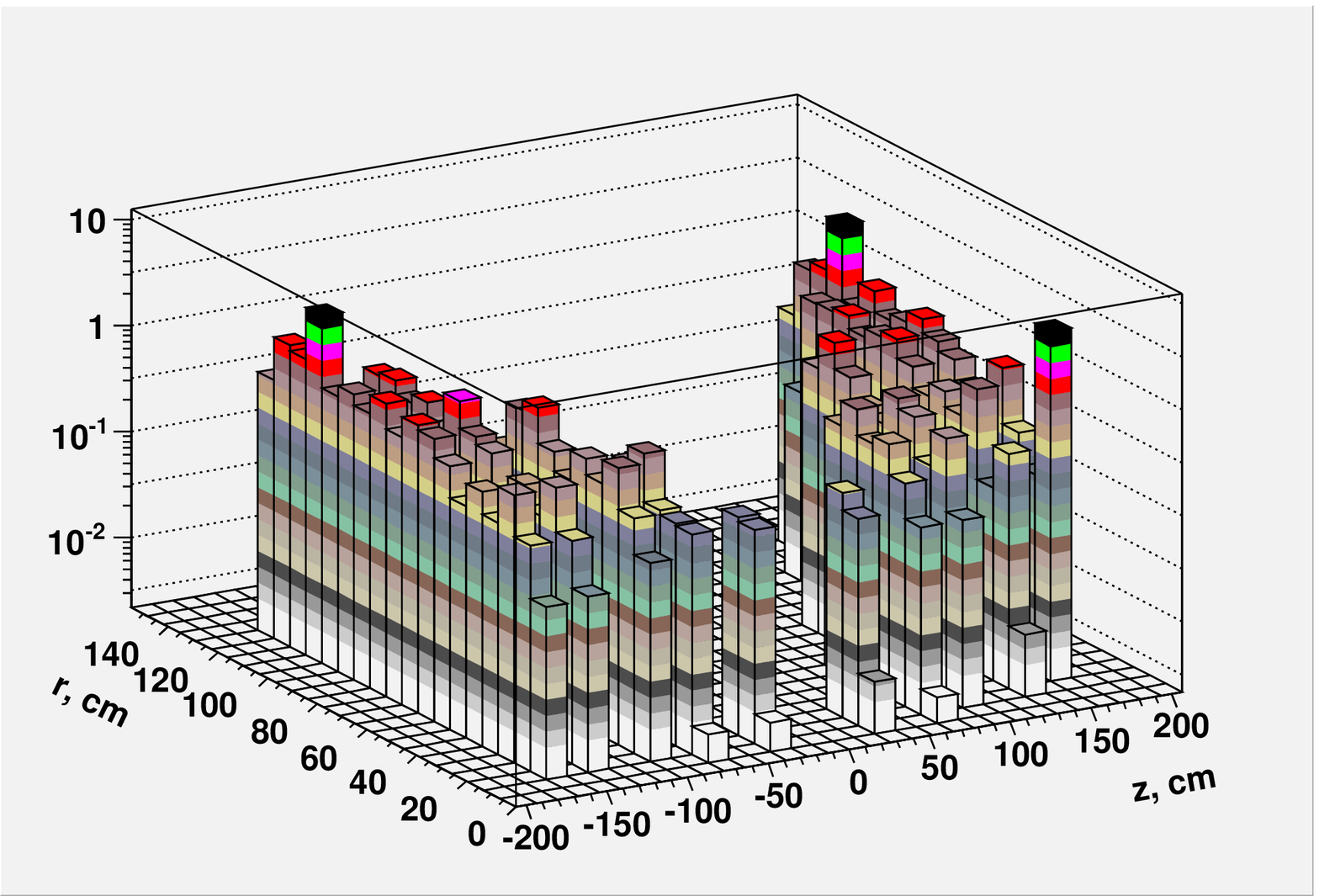,width=\linewidth}
\caption{RZ distribution of hits per bunch in Tracker Endcap. No spoilers. Background is created by particles coming from positron tunnel only. }
\end{minipage}
\end{figure}

\newpage
\begin{figure}[hbt!]
\begin{minipage}[b]{0.90\linewidth}
\centering\epsfig{figure=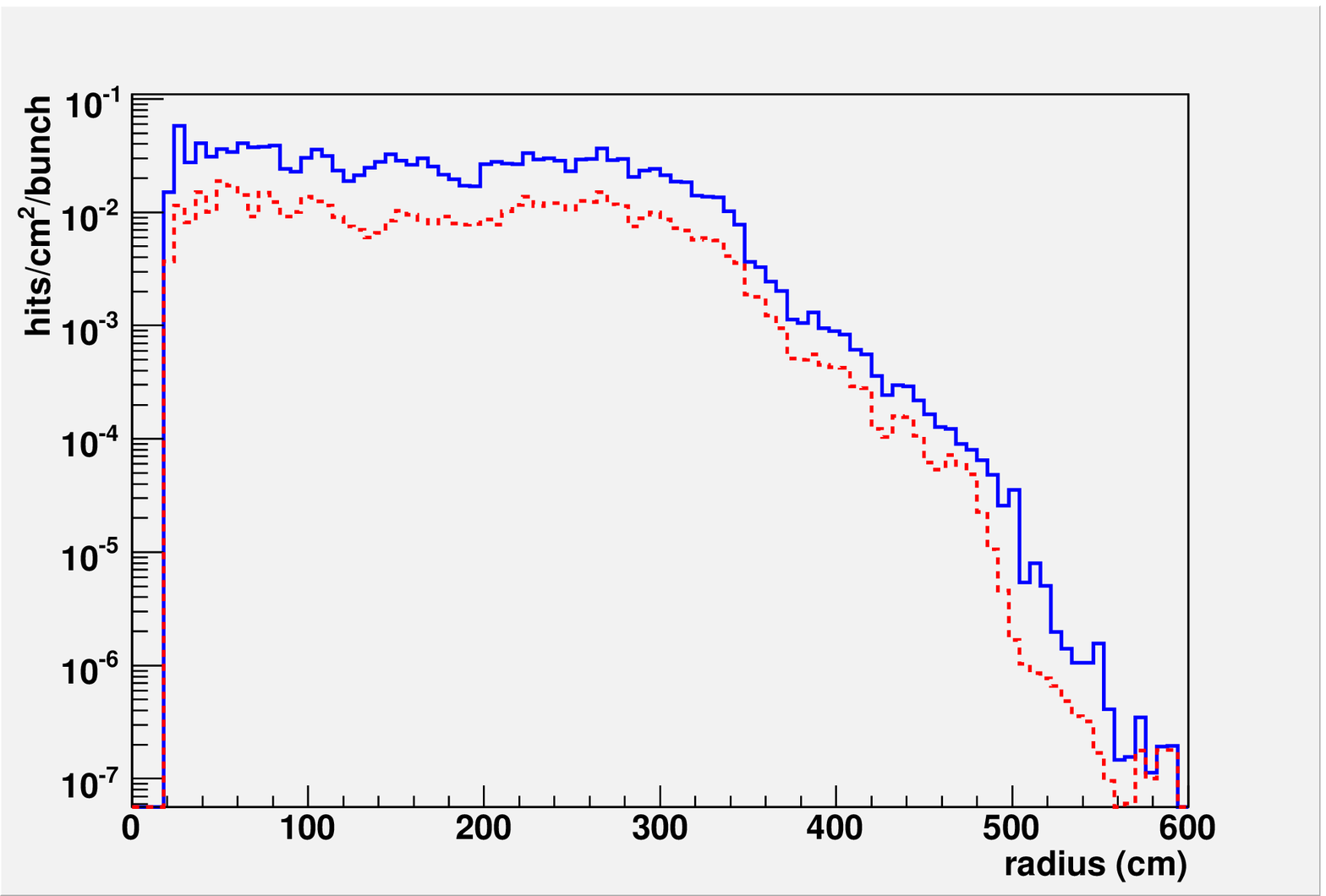,width=\linewidth}
\caption{Radial distribution of hits in Muon Endcap. Solid line - total background, dashed line -
background after bunch crossing. No spoilers. Particles coming from electron and positron tunnels are included.}
\end{minipage}

\begin{minipage}[b]{1.0\linewidth}
\vspace{-1.5cm}
\centering\epsfig{figure=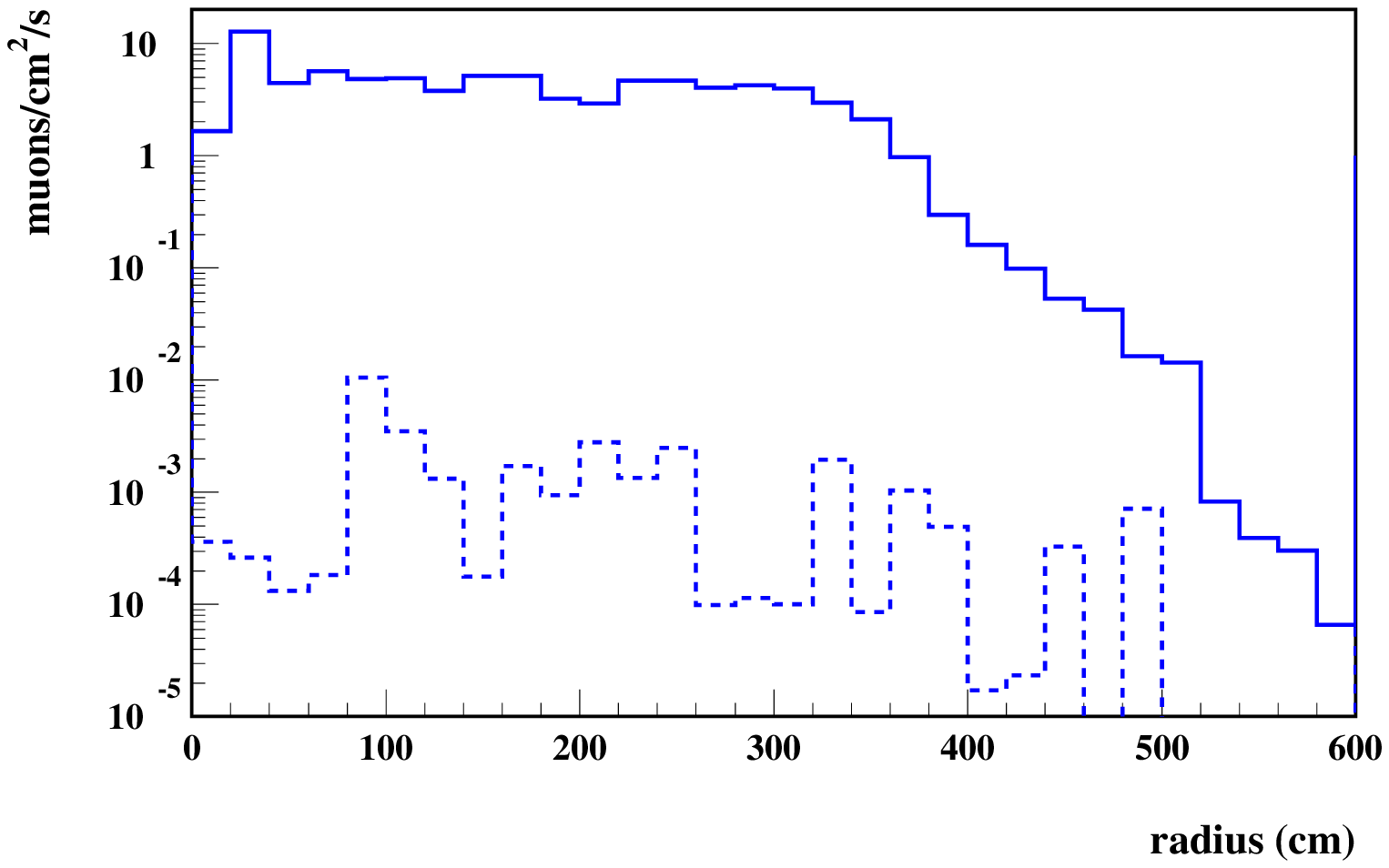,width=\linewidth}
\vspace{-2.0cm}
\caption{Radial distribution of muons at the Muon Endcap entrance. 
Solid line - no spoilers, 
dashed line - tunnel
 with spoilers.
Muons coming from electron and positron tunnels are included.}
\end{minipage}
\end{figure}

\newpage
\begin{figure}[hbt!]
\begin{minipage}[b]{0.91\linewidth}
\centering\epsfig{figure=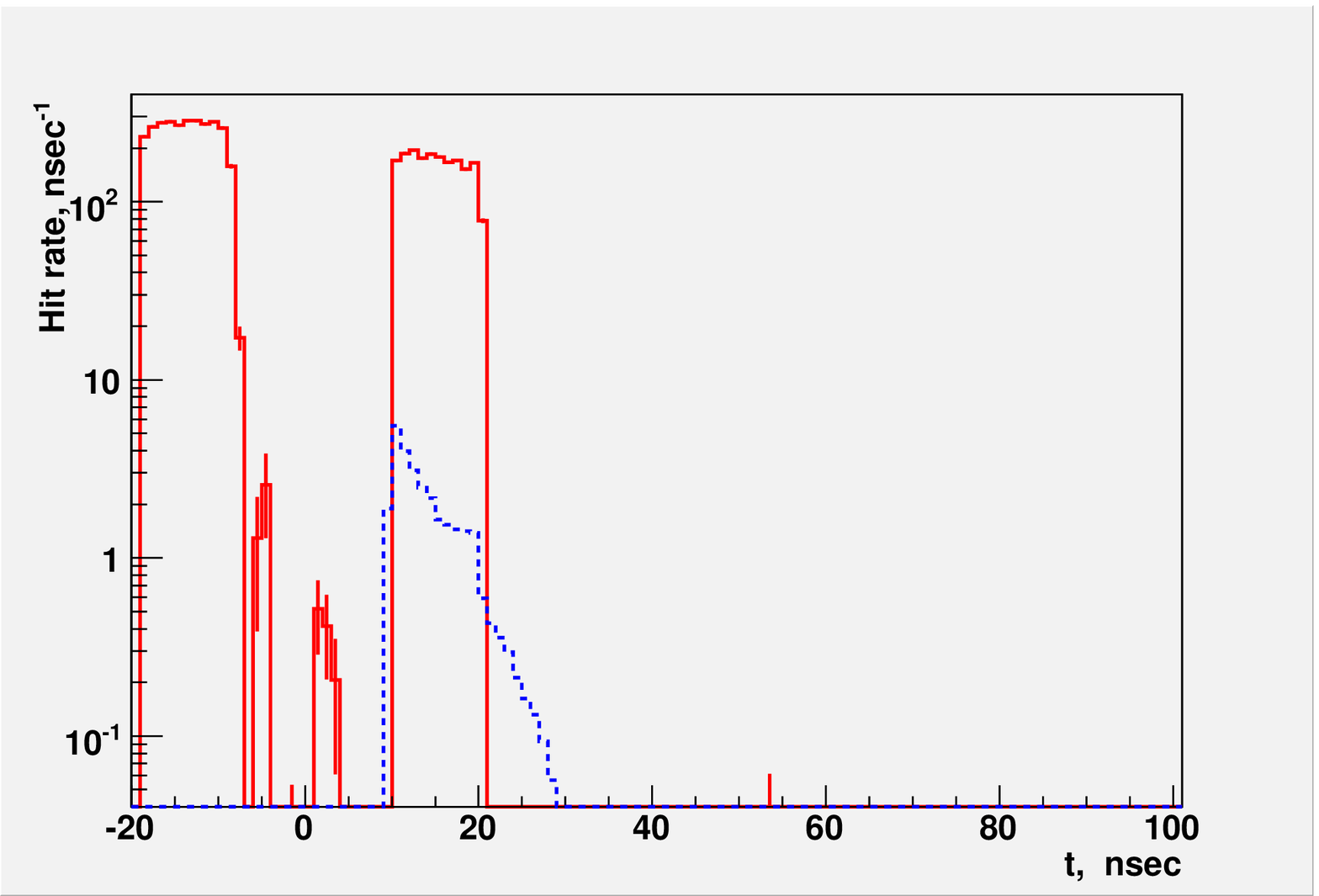,width=\linewidth}
\caption{Time distribution of hit rates in Muon Endcap. 
Solid line - BDS background (no spoilers), dashed line - $e^+ e^-$ events. 
BDS background is from positron tunnel only. }
\end{minipage}
\begin{minipage}[b]{0.91\linewidth}
\centering\epsfig{figure=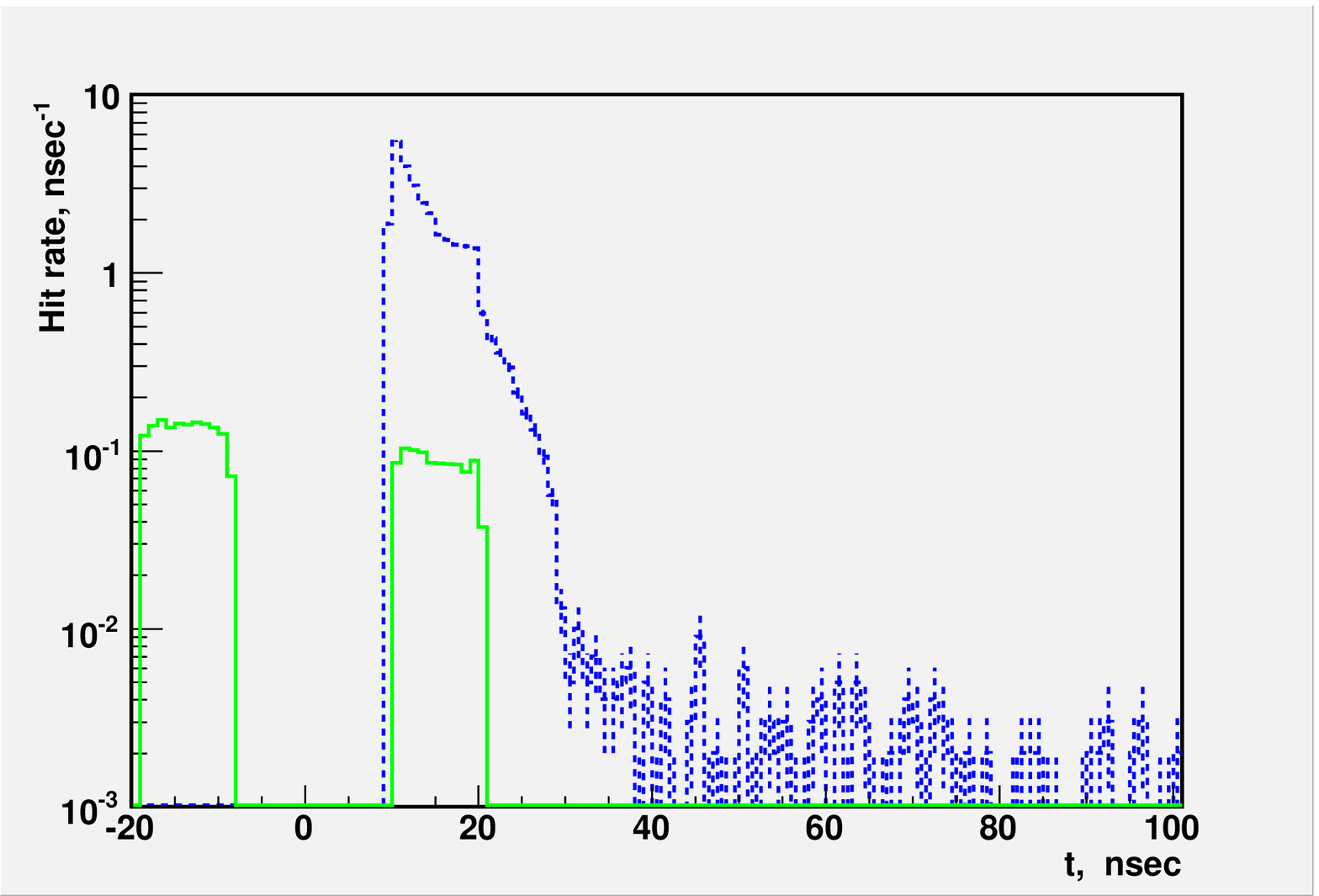,width=\linewidth}
\caption{Time distribution of hit rates in Muon Endcap. 
Solid line - BDS background (with spoilers), dashed line - $e^+ e^-$ events.
BDS background is from positron tunnel only. }
\end{minipage}
\end{figure}

\begin{figure}[hbt!]
\begin{minipage}[b]{0.91\linewidth}
\centering\epsfig{figure=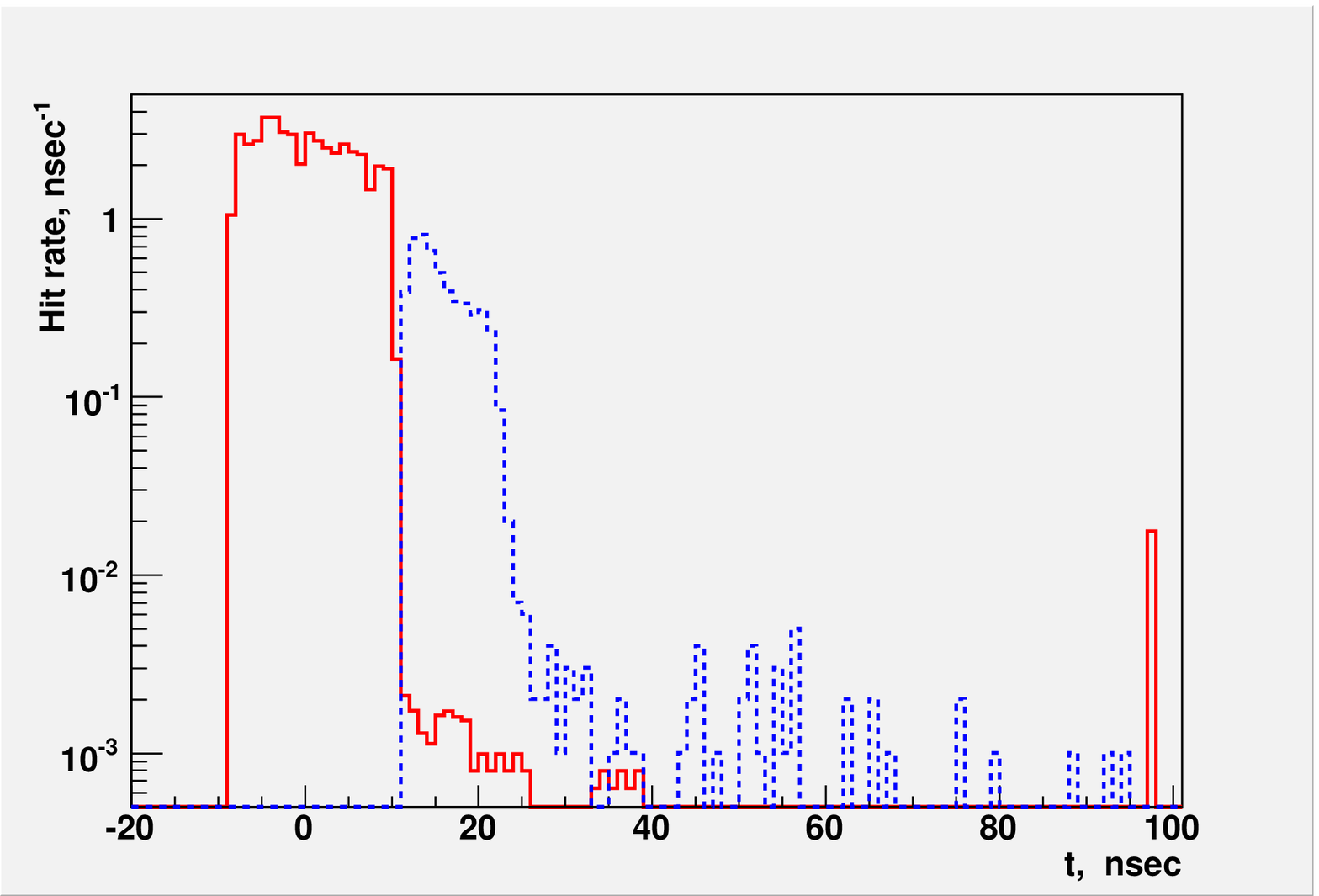,width=\linewidth}
\caption{Time distribution of hit rates in Muon Barrel. 
Solid line - BDS background (no spoilers), dashed line - $e^+ e^-$ events.
BDS background is from positron tunnel only. }
\end{minipage}
\begin{minipage}[b]{0.91\linewidth}
\centering\epsfig{figure=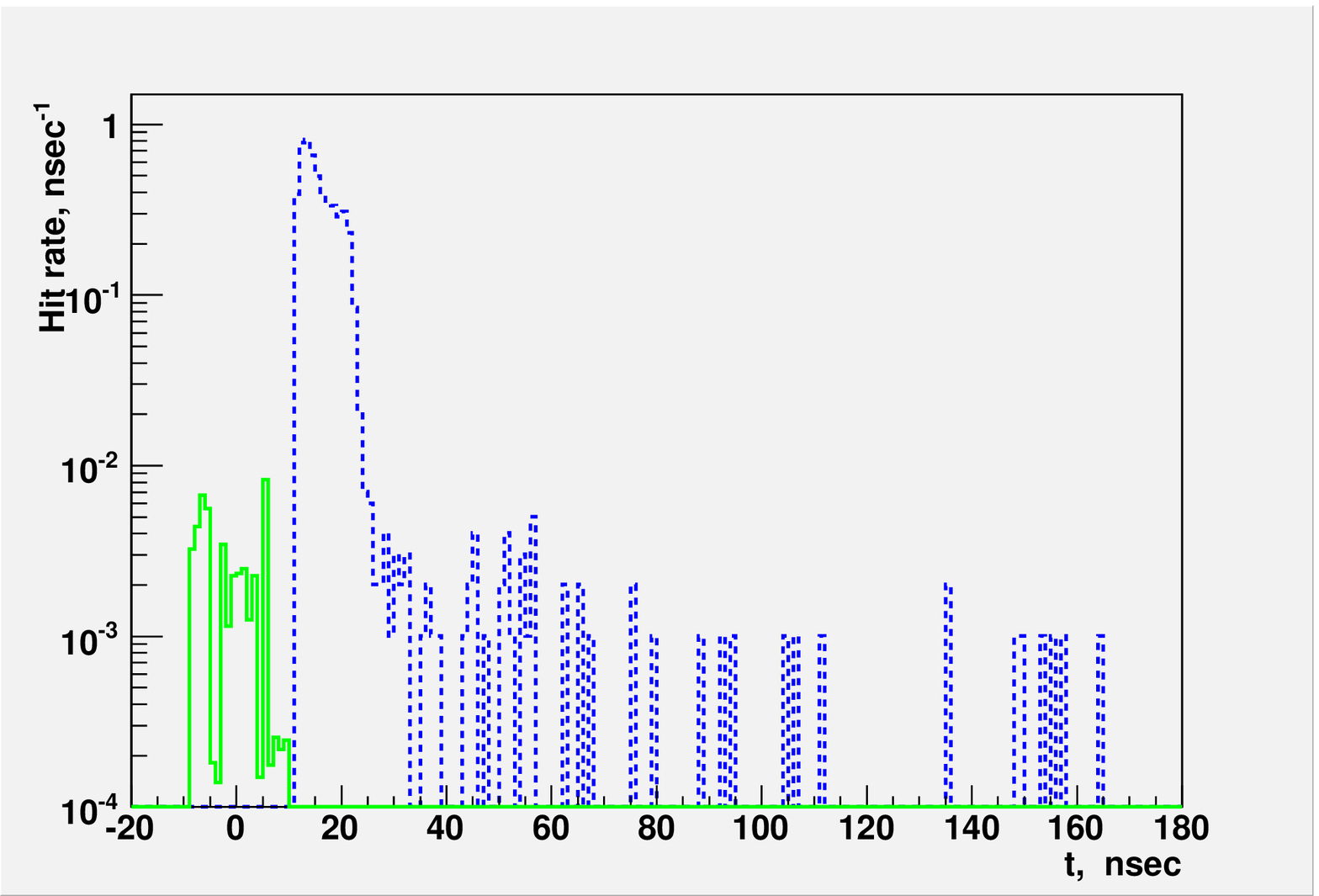,width=\linewidth}
\caption{Time distribution of hit rates in Muon Barrel. 
Solid line - BDS background (with spoilers), dashed line - $e^+ e^-$ events.
BDS background is from positron tunnel only. }
\end{minipage}
\end{figure}

\newpage
\begin{figure}[hbt!]
\begin{minipage}[b]{0.91\linewidth}
\centering\epsfig{figure=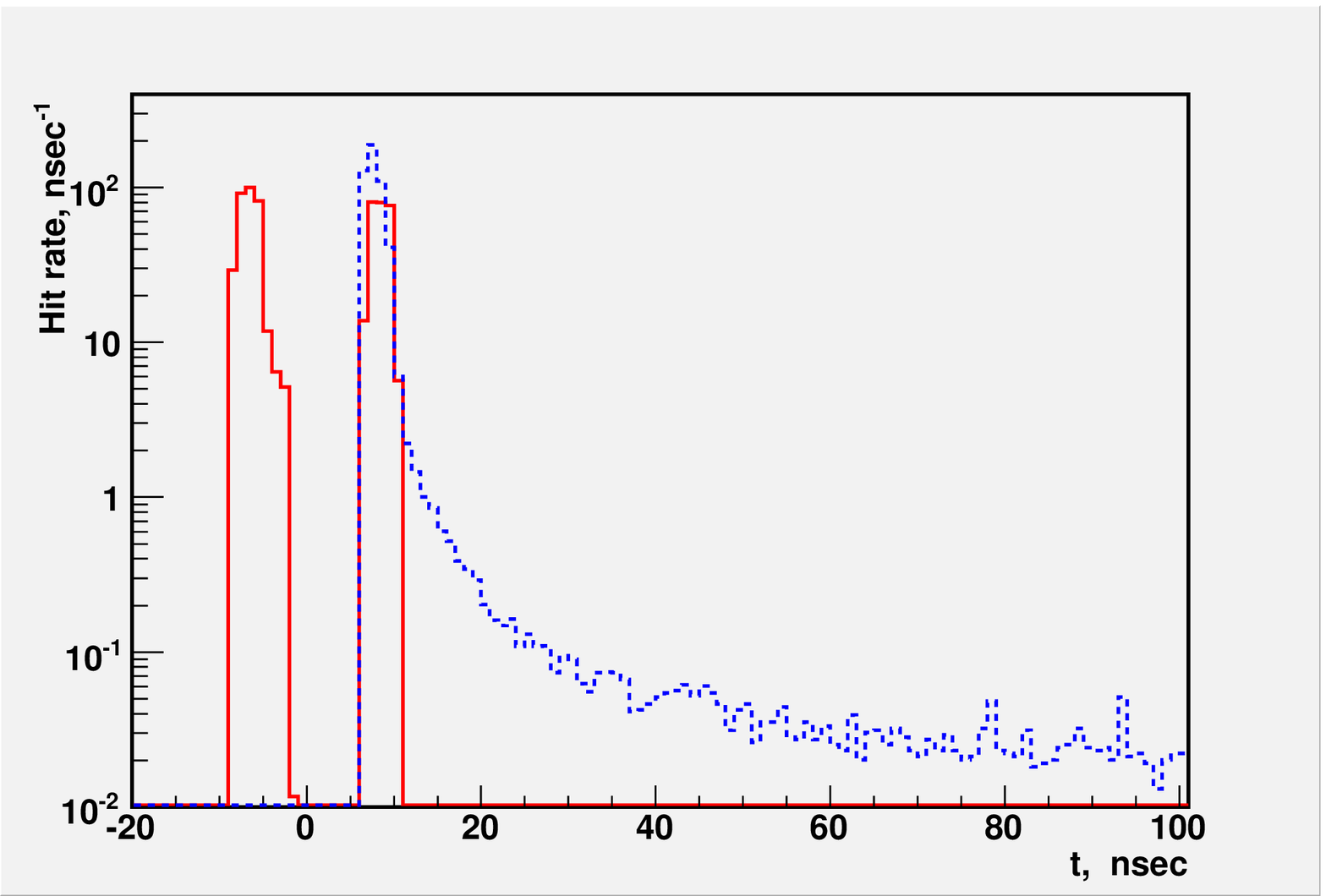,width=\linewidth}
\caption{Time distribution of hit rates in Hcal Endcap.
Solid line - BDS background (no spoilers), dashed line - $e^+ e^-$ events.
BDS background is from positron tunnel only. }
\end{minipage}
\begin{minipage}[b]{0.91\linewidth}
\centering\epsfig{figure=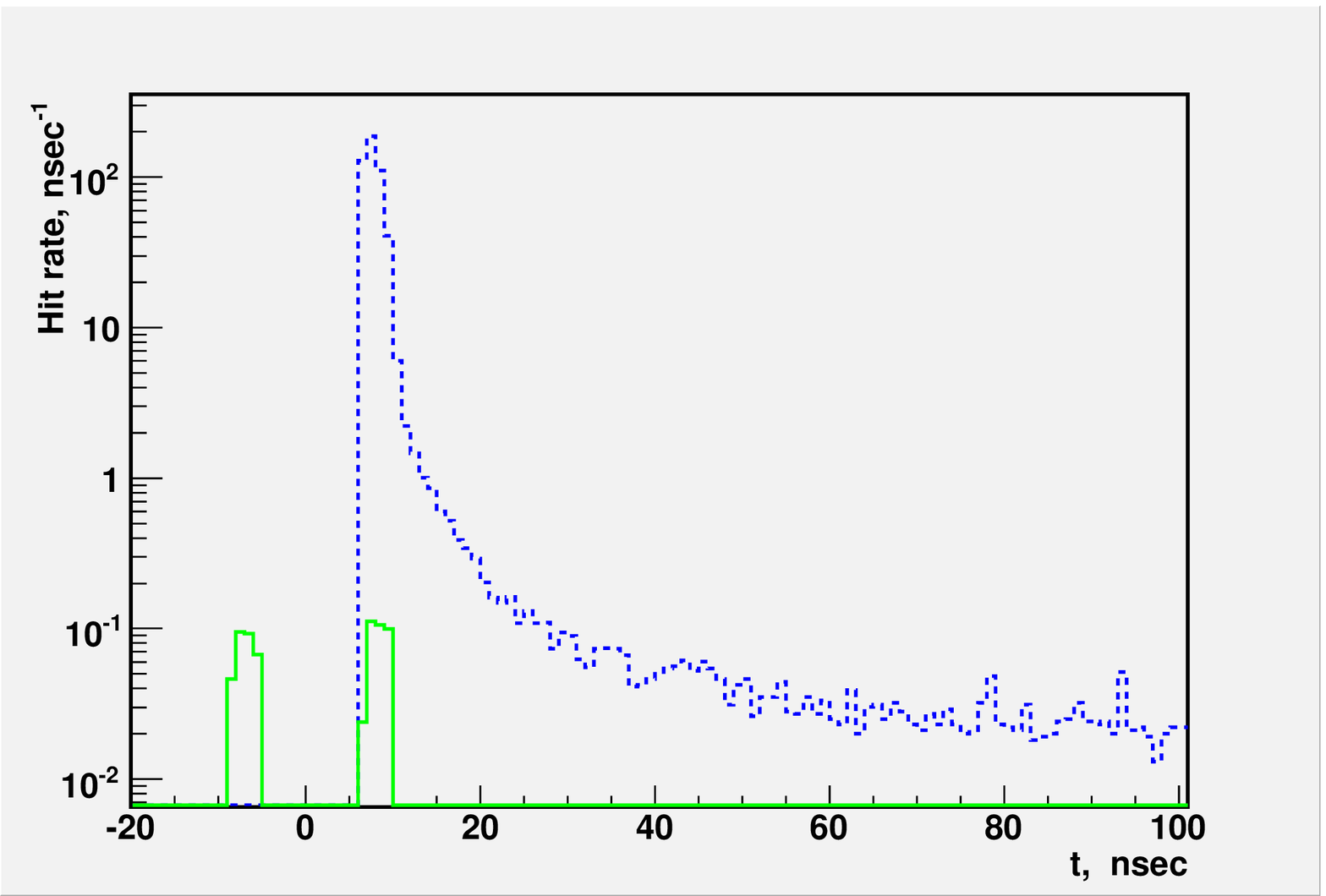,width=\linewidth}
\caption{Time distribution of hit rates in Hcal Endcap. 
Solid line - BDS background (with spoilers), dashed line - $e^+ e^-$ events.
BDS background is from positron tunnel only. }
\end{minipage}
\end{figure}

\newpage
\begin{figure}[hbt!]
\begin{minipage}[b]{0.91\linewidth}
\centering\epsfig{figure=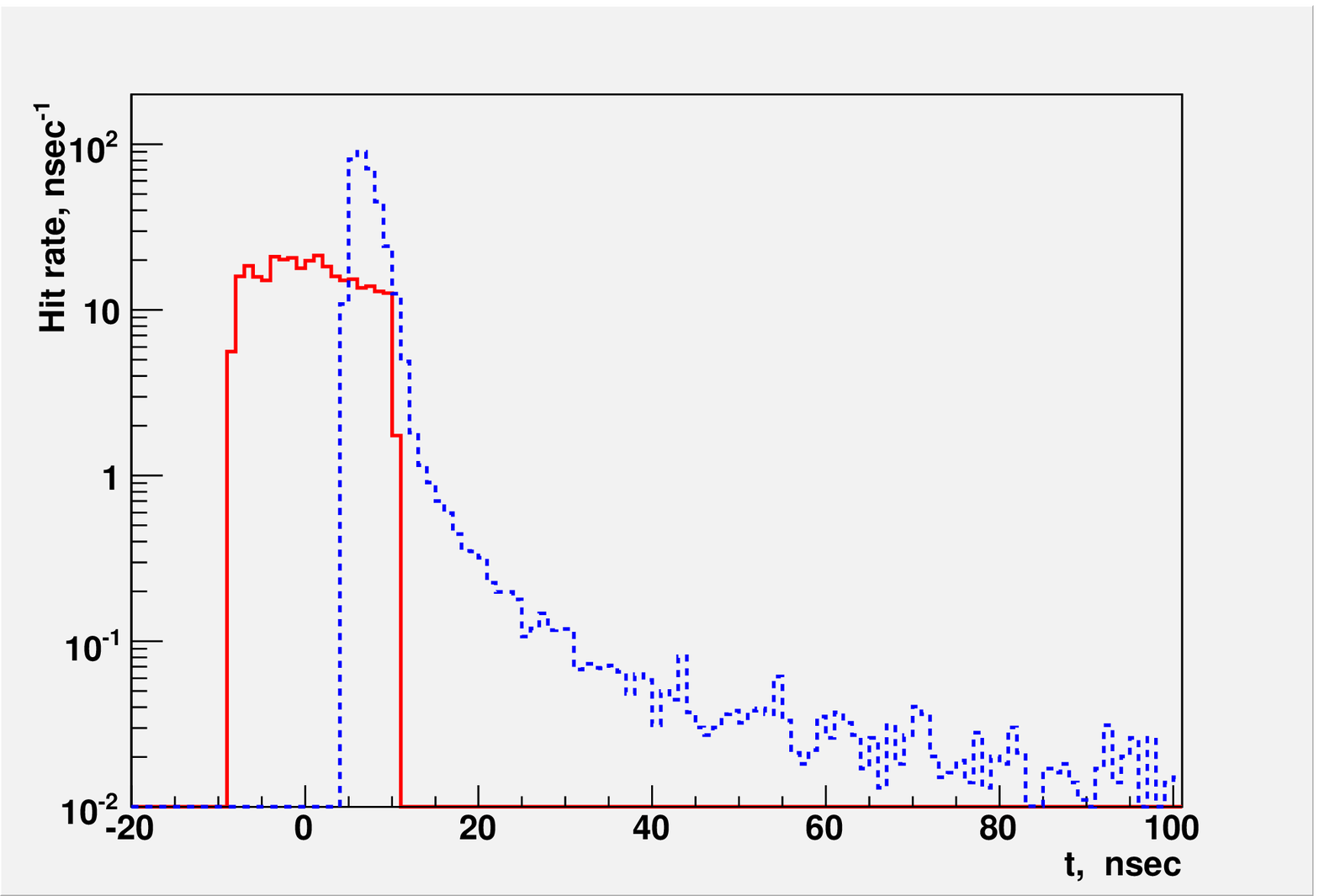,width=\linewidth}
\caption{Time distribution of hit rates in Hcal Barrel.
Solid line - BDS background (no spoilers), dashed line - $e^+ e^-$ events.
BDS background is from positron tunnel only. }
\end{minipage}
\begin{minipage}[b]{0.91\linewidth}
\centering\epsfig{figure=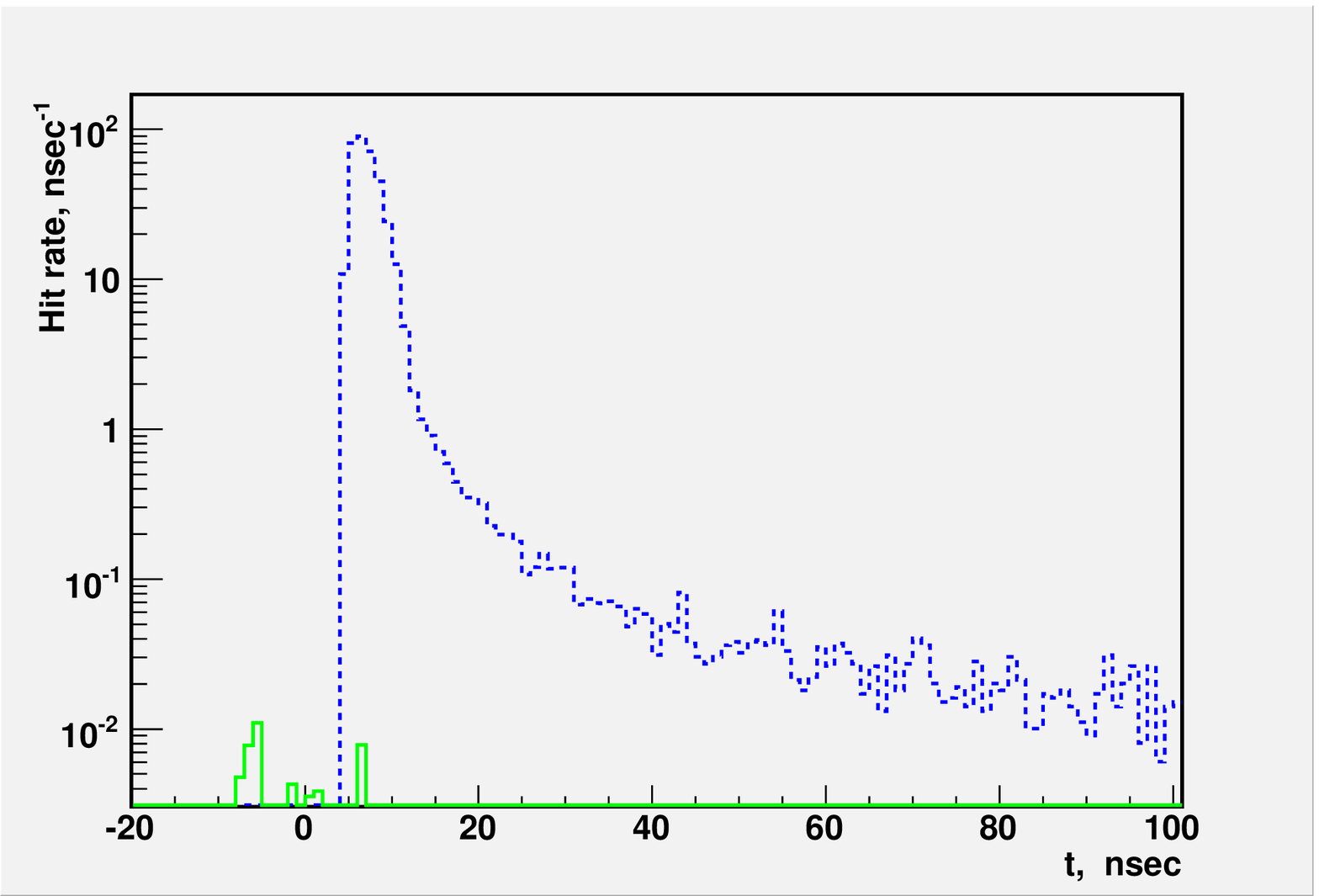,width=\linewidth}
\caption{Time distribution of hit rates in Hcal Barrel. 
Solid line - BDS background (with spoilers), dashed line - $e^+ e^-$ events.
BDS background is from positron tunnel only. }
\end{minipage}
\end{figure}

\newpage
\begin{figure}[hbt!]
\begin{minipage}[b]{0.91\linewidth}
\centering\epsfig{figure=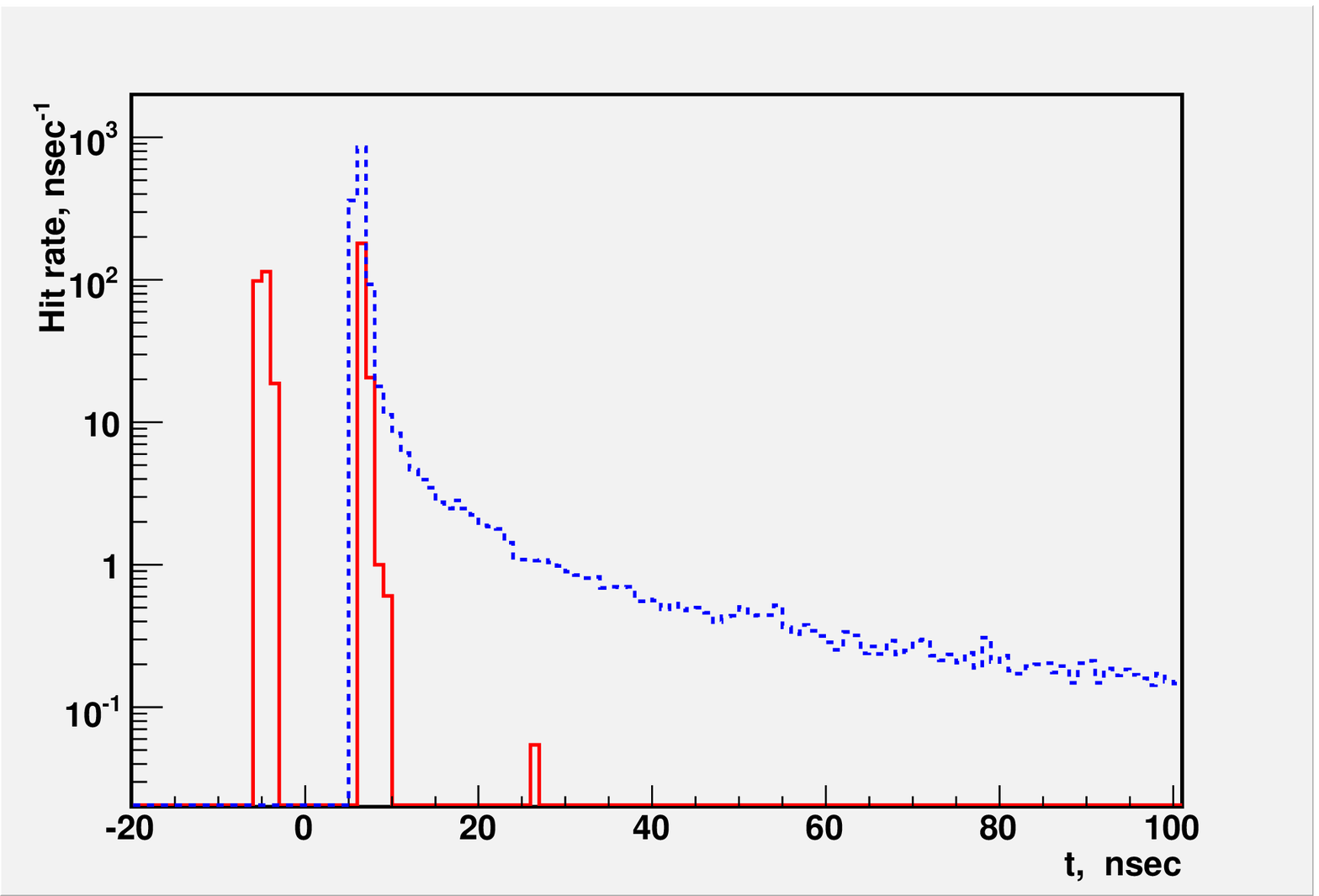,width=\linewidth}
\caption{Time distribution of hit rates in Ecal Endcap. 
Solid line - BDS background (no spoilers), dashed line - $e^+ e^-$ events.
BDS background is from positron tunnel only. }
\end{minipage}
\begin{minipage}[b]{0.91\linewidth}
\centering\epsfig{figure=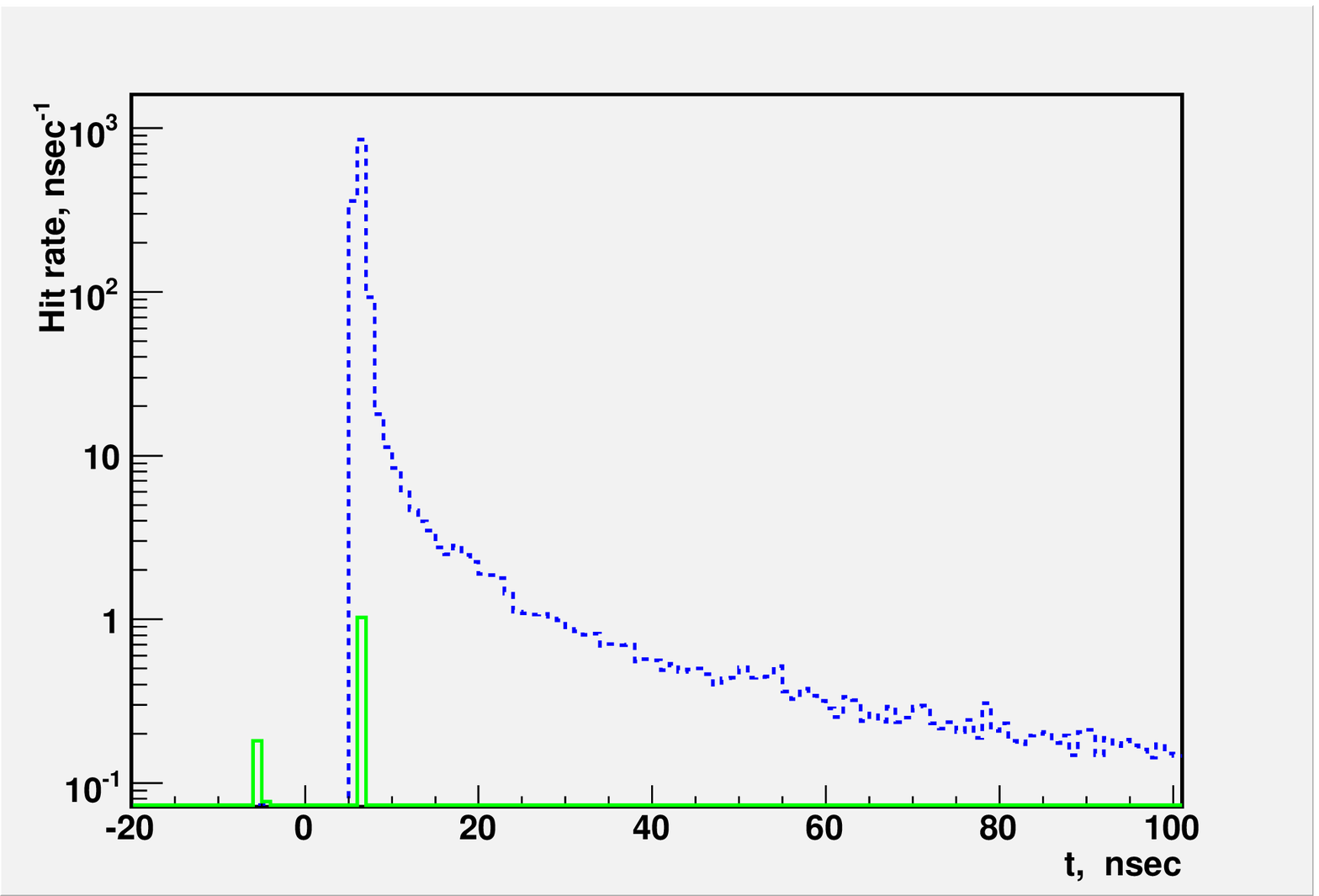,width=\linewidth}
\caption{Time distribution of hit rates in Ecal Endcap. 
Solid line - BDS background (with spoilers), dashed line - $e^+ e^-$ events.
BDS background is from positron tunnel only. }
\end{minipage}
\end{figure}

\newpage
\begin{figure}[hbt!]
\begin{minipage}[b]{0.91\linewidth}
\centering\epsfig{figure=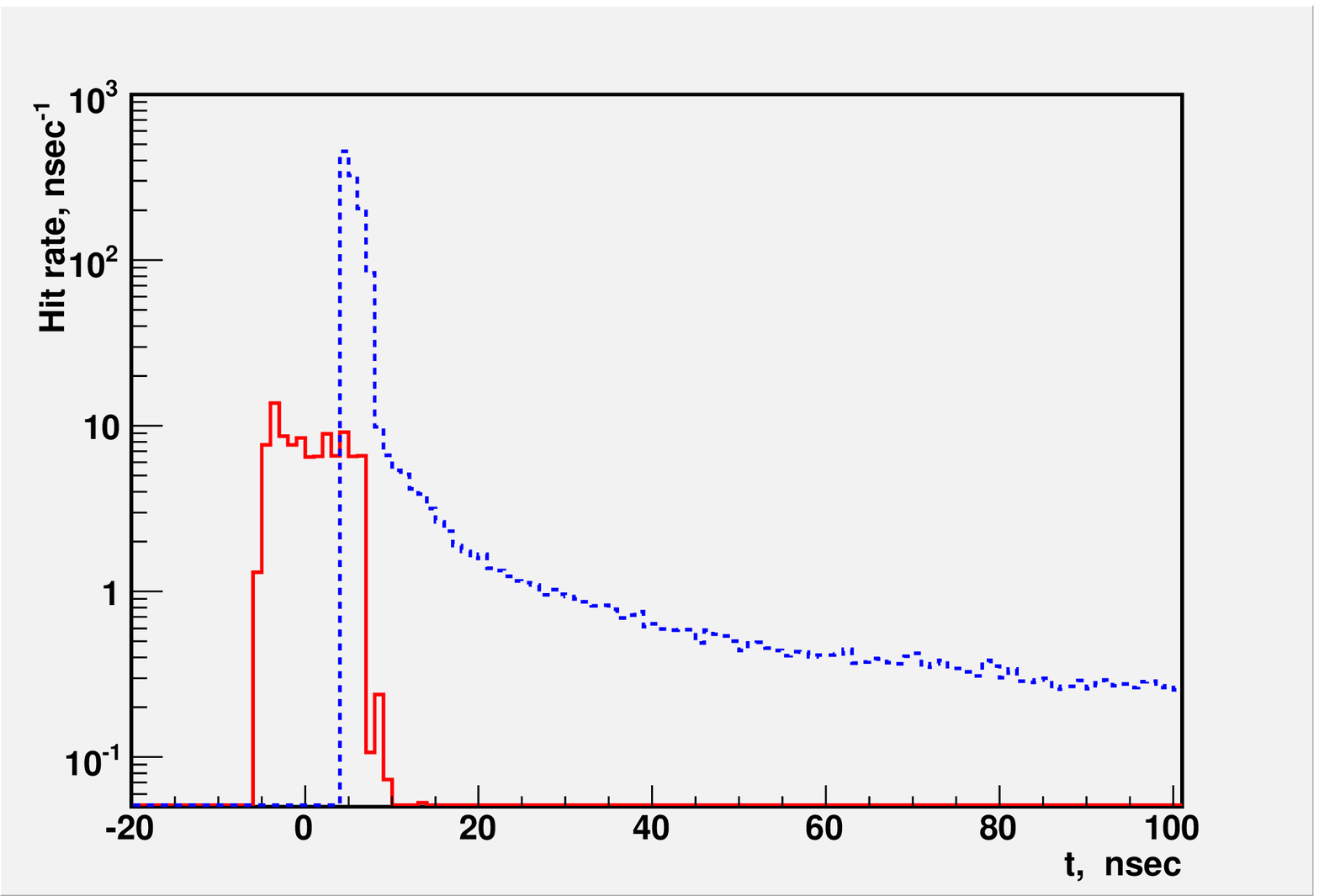,width=\linewidth}
\caption{Time distribution of hit rates in Ecal Barrel. 
Solid line - BDS background (no spoilers), dashed line - $e^+ e^-$ events.
BDS background is from positron tunnel only. }
\end{minipage}
\begin{minipage}[b]{0.91\linewidth}
\centering\epsfig{figure=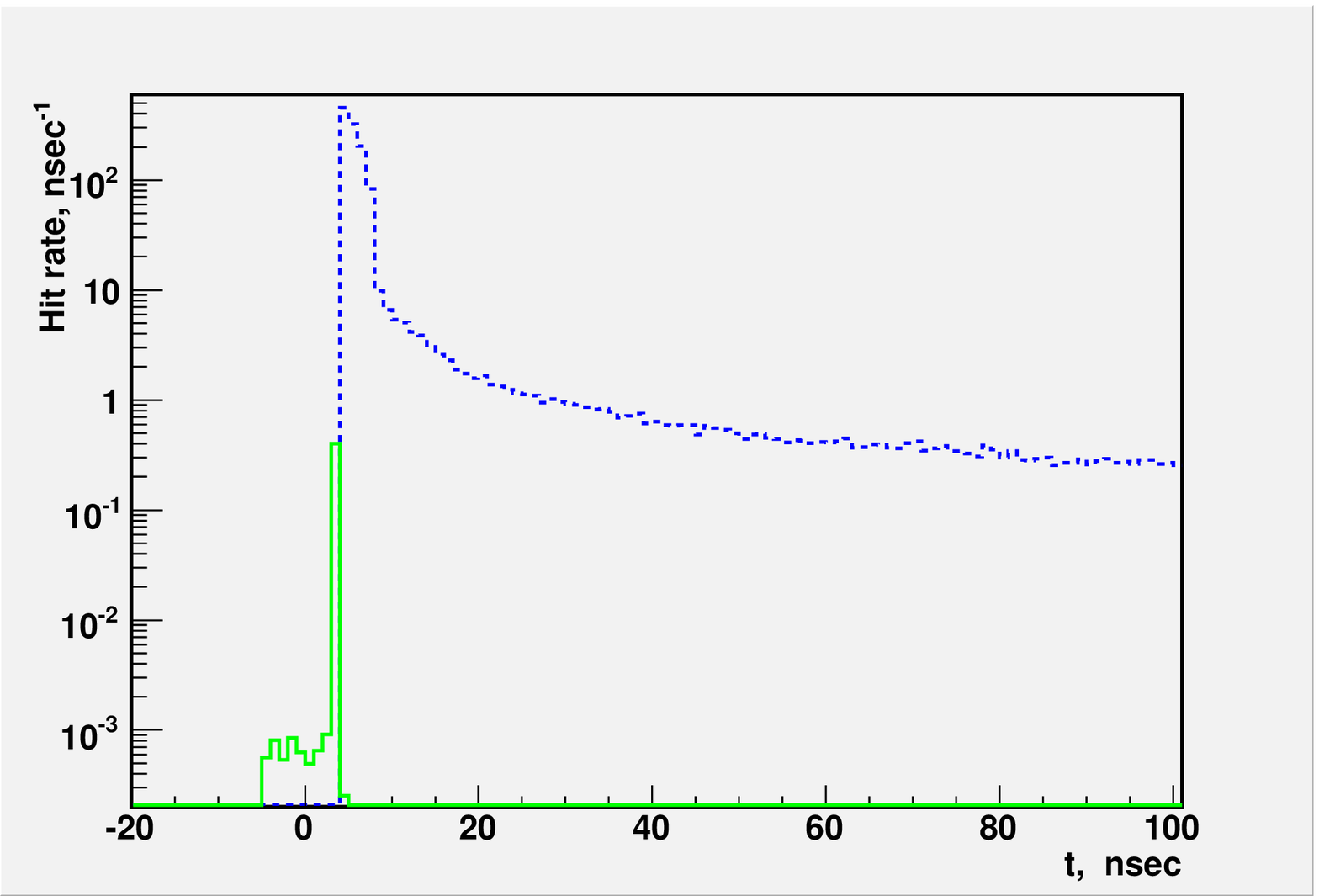,width=\linewidth}
\caption{Time distribution of hit rates in Ecal Barrel. 
Solid line - BDS background (with spoilers), dashed line - $e^+ e^-$ events.
BDS background is from positron tunnel only. }
\end{minipage}
\end{figure}

\newpage
\begin{figure}[hbt!]
\begin{minipage}[b]{0.91\linewidth}
\centering\epsfig{figure=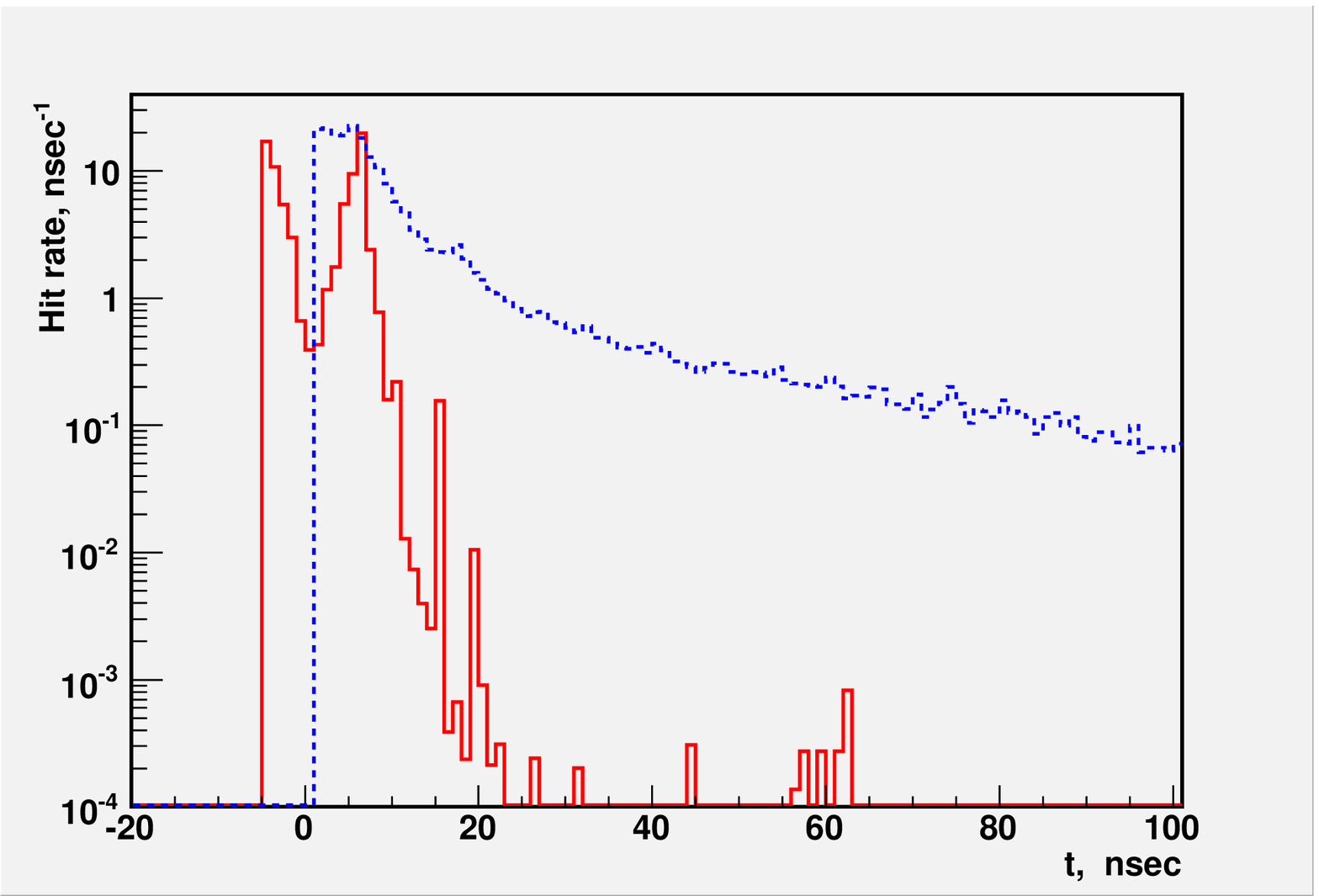,width=\linewidth}
\caption{Time distribution of hit rates in Tracker Endcap. 
Solid line - BDS background (no spoilers), dashed line - $e^+ e^-$ events.
BDS background is from positron tunnel only. }
\end{minipage}
\begin{minipage}[b]{0.91\linewidth}
\centering\epsfig{figure=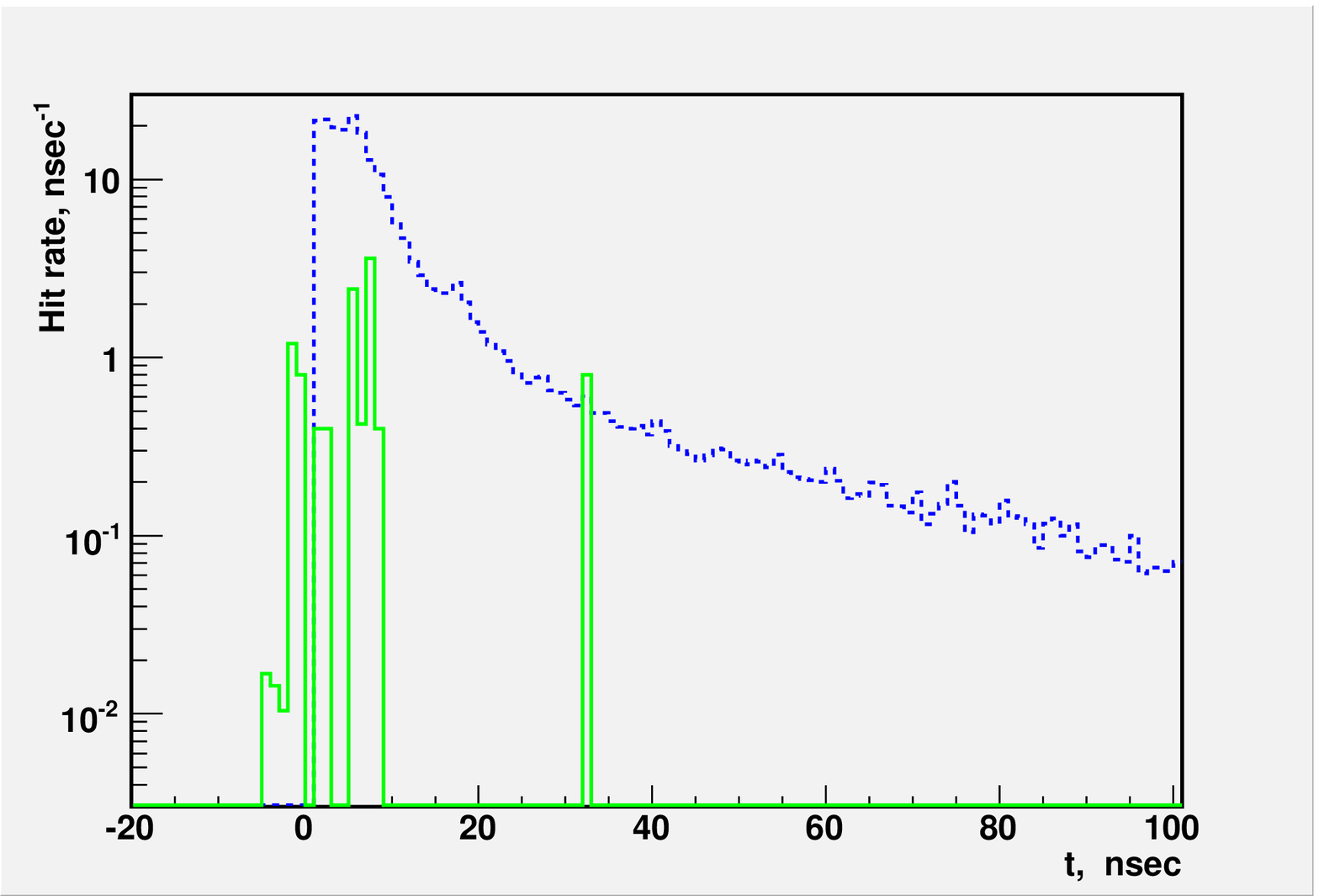,width=\linewidth}
\caption{Time distribution of hit rates in Tracker Endcap.  
Solid line - BDS background (with spoilers), dashed line - $e^+ e^-$ events.
BDS background is from positron tunnel only. }
\end{minipage}
\end{figure}

\newpage
\begin{figure}[hbt!]
\begin{minipage}[b]{0.91\linewidth}
\centering\epsfig{figure=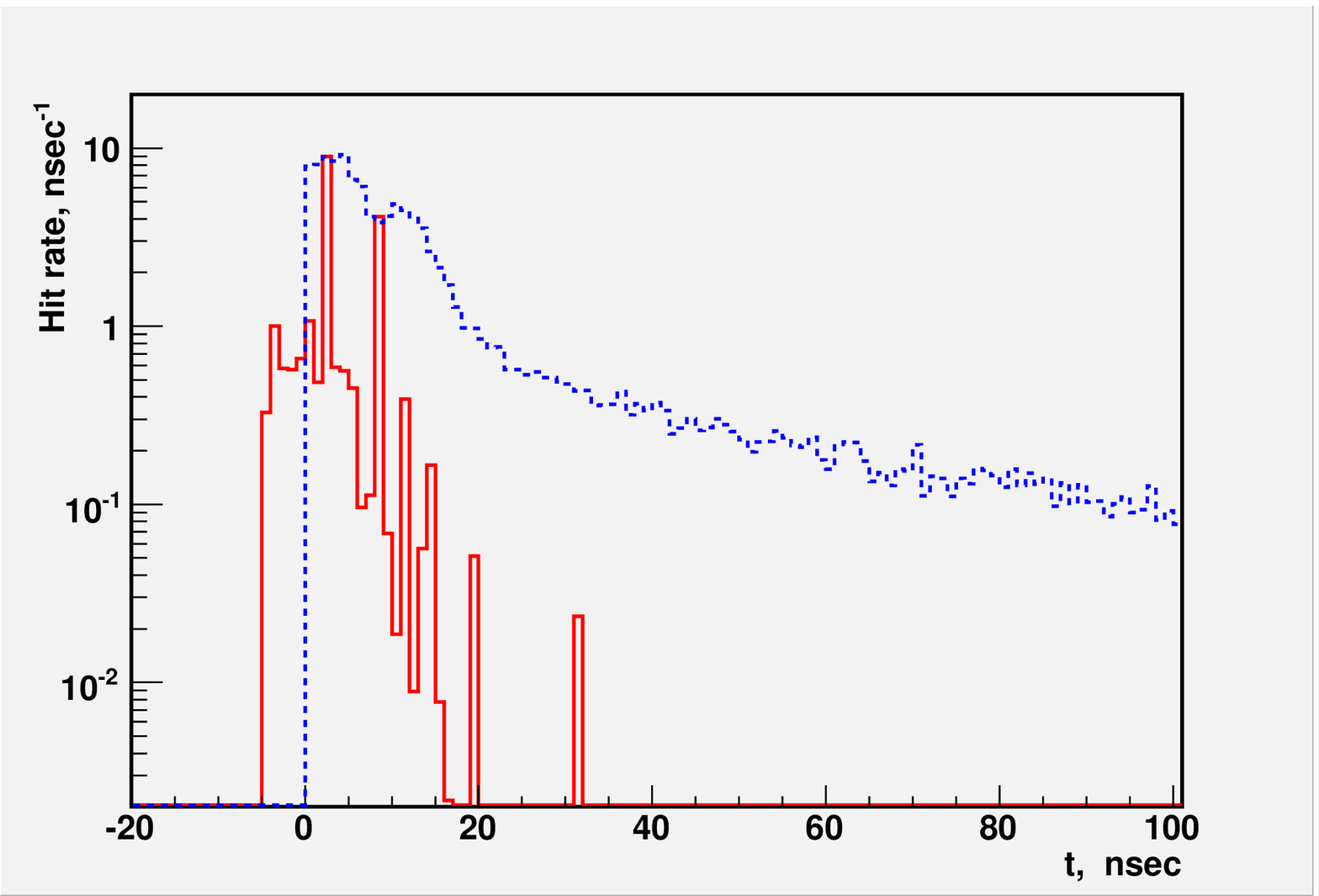,width=\linewidth}
\caption{Time distribution of hit rates in Tracker Barrel.
Solid line - BDS background (no spoilers), dashed line - $e^+ e^-$ events.
BDS background is from positron tunnel only. }
\end{minipage}
\begin{minipage}[b]{0.91\linewidth}
\centering\epsfig{figure=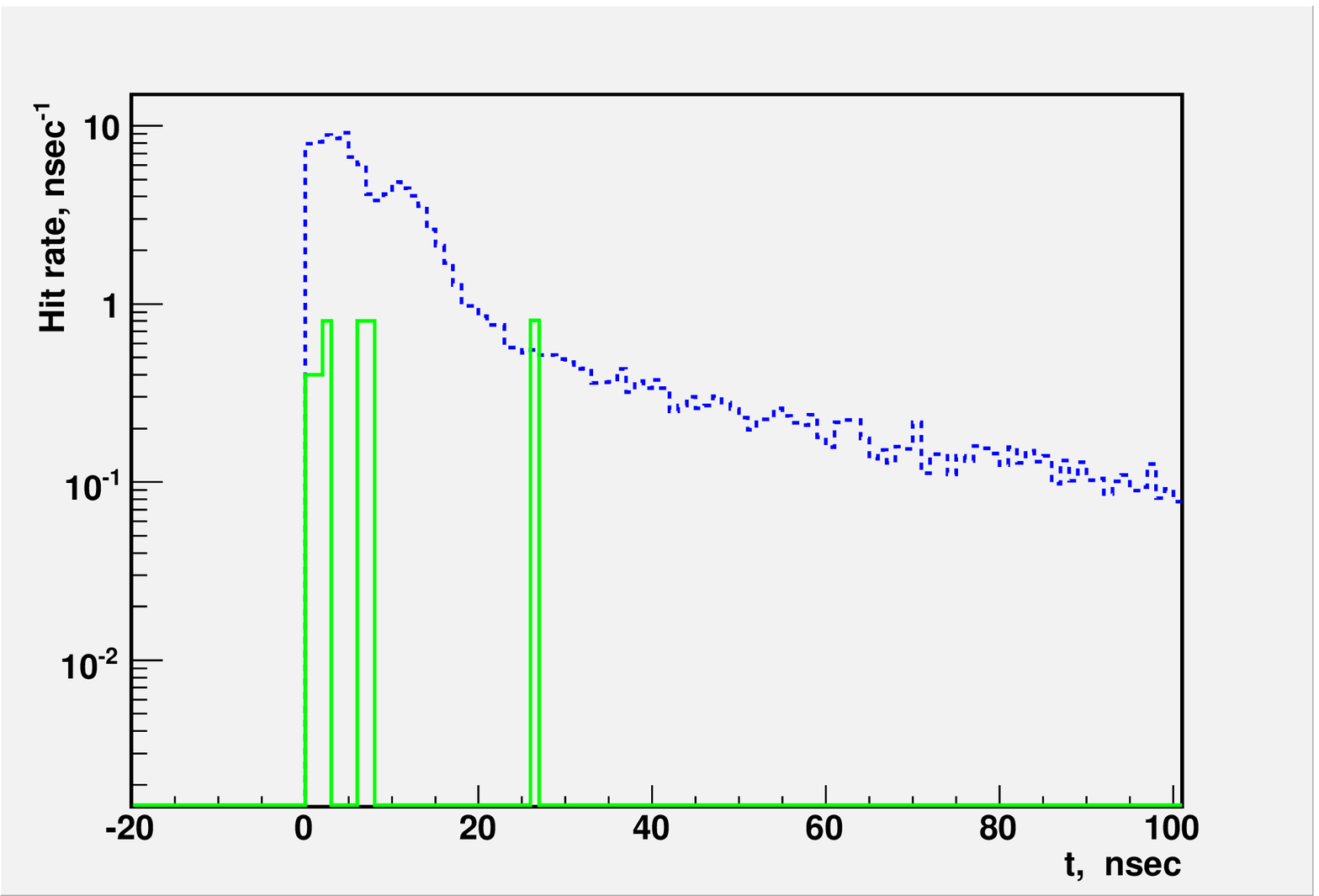,width=\linewidth}
\caption{Time distribution of hit rates in Tracker Barrel. 
Solid line - BDS background (with spoilers), dashed line - $e^+ e^-$ events.
BDS background is from positron tunnel only. }
\end{minipage}
\end{figure}

\clearpage
\begin{figure}[hbt!]
\begin{minipage}[b]{0.91\linewidth}
\centering\epsfig{figure=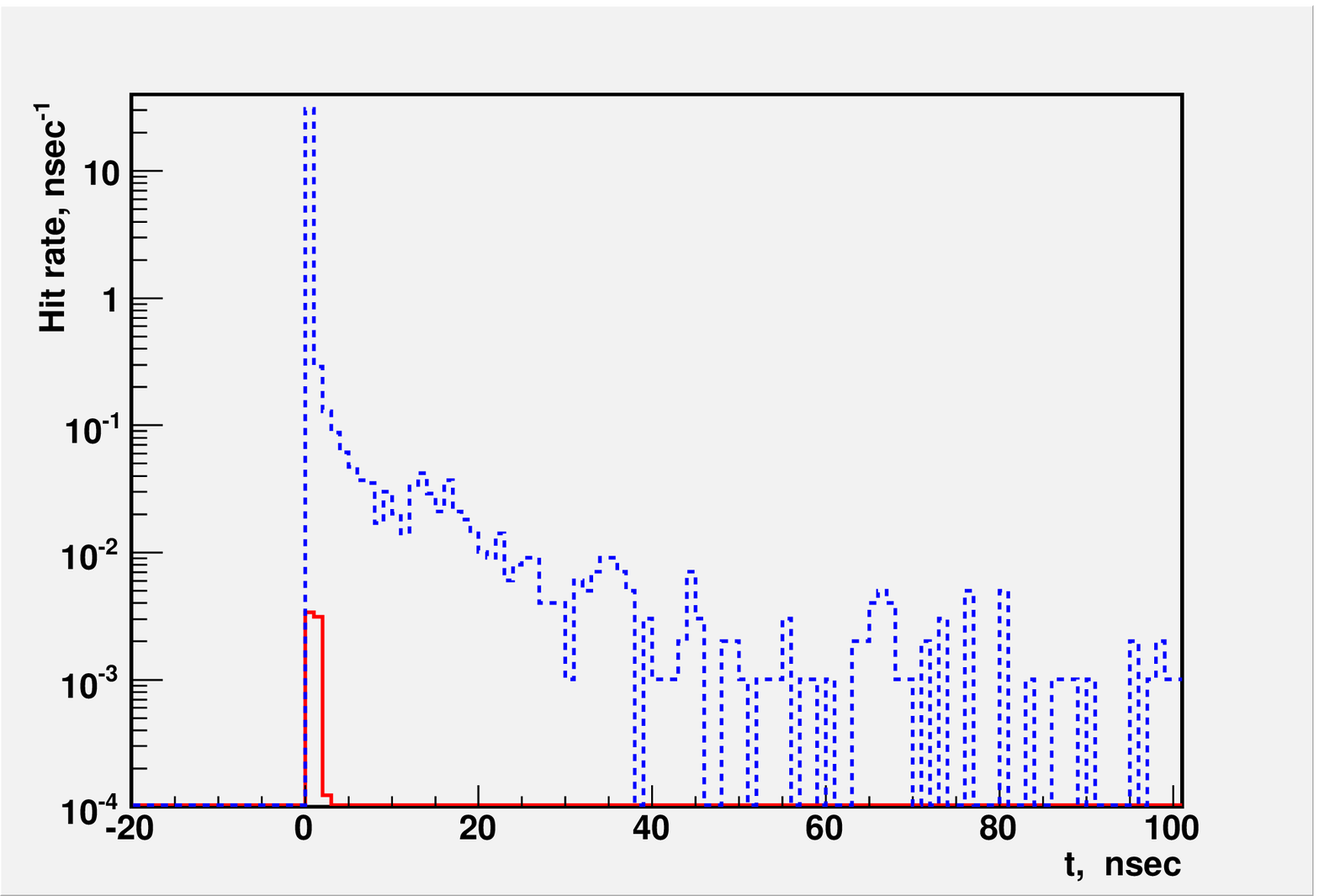,width=\linewidth}
\caption{Time distribution of hit rates in Vertex Endcap.
Solid line - BDS background (no spoilers), dashed line - $e^+ e^-$ events.
BDS background is from positron tunnel only. }
\end{minipage}
\begin{minipage}[b]{0.91\linewidth}
\centering\epsfig{figure=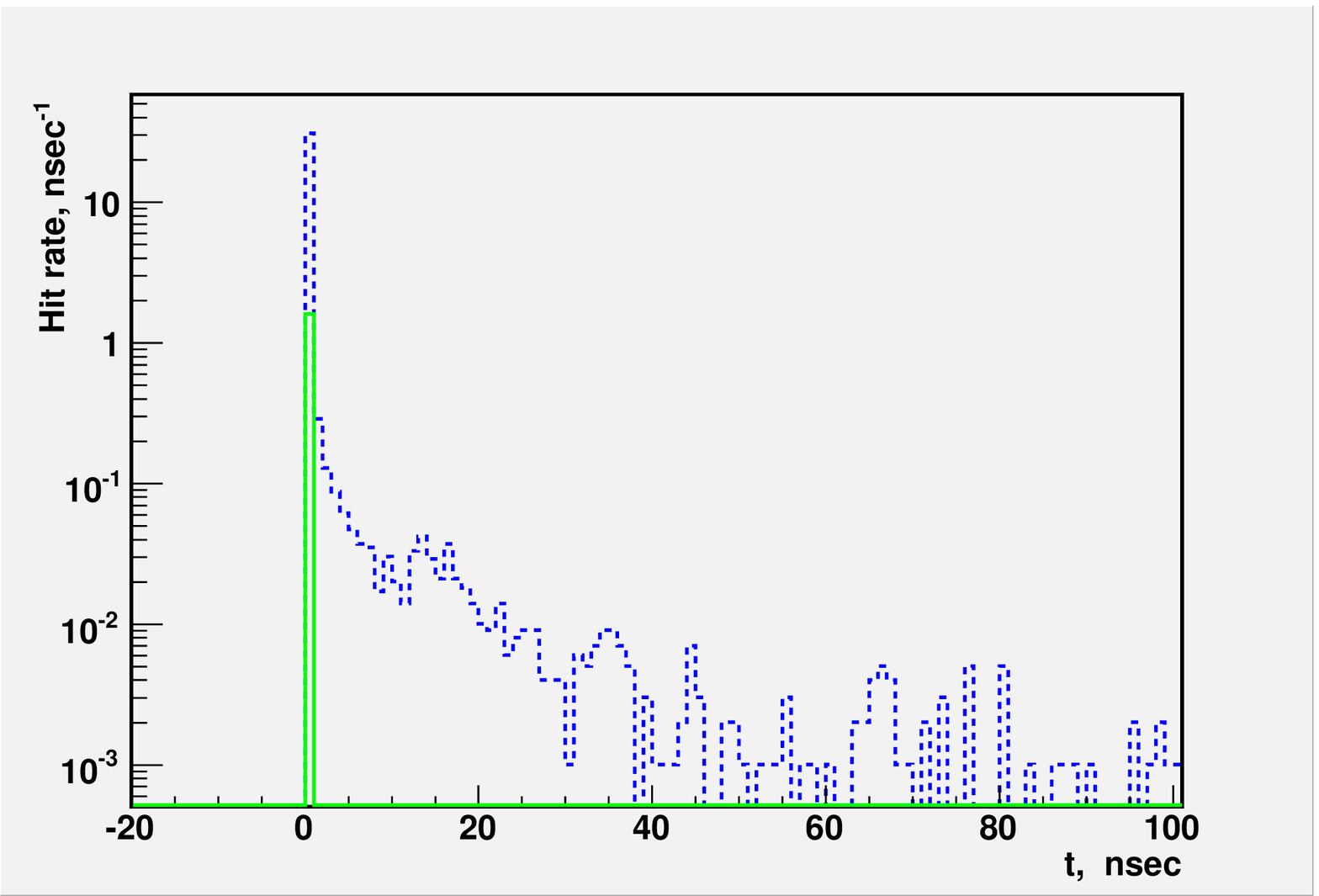,width=\linewidth}
\caption{Time distribution of hit rates in Vertex Endcap. 
Solid line - BDS background (with spoilers), dashed line - $e^+ e^-$ events.
BDS background is from positron tunnel only. }
\end{minipage}
\end{figure}

\newpage
\begin{figure}[hbt!]
\begin{minipage}[b]{0.91\linewidth}
\centering\epsfig{figure=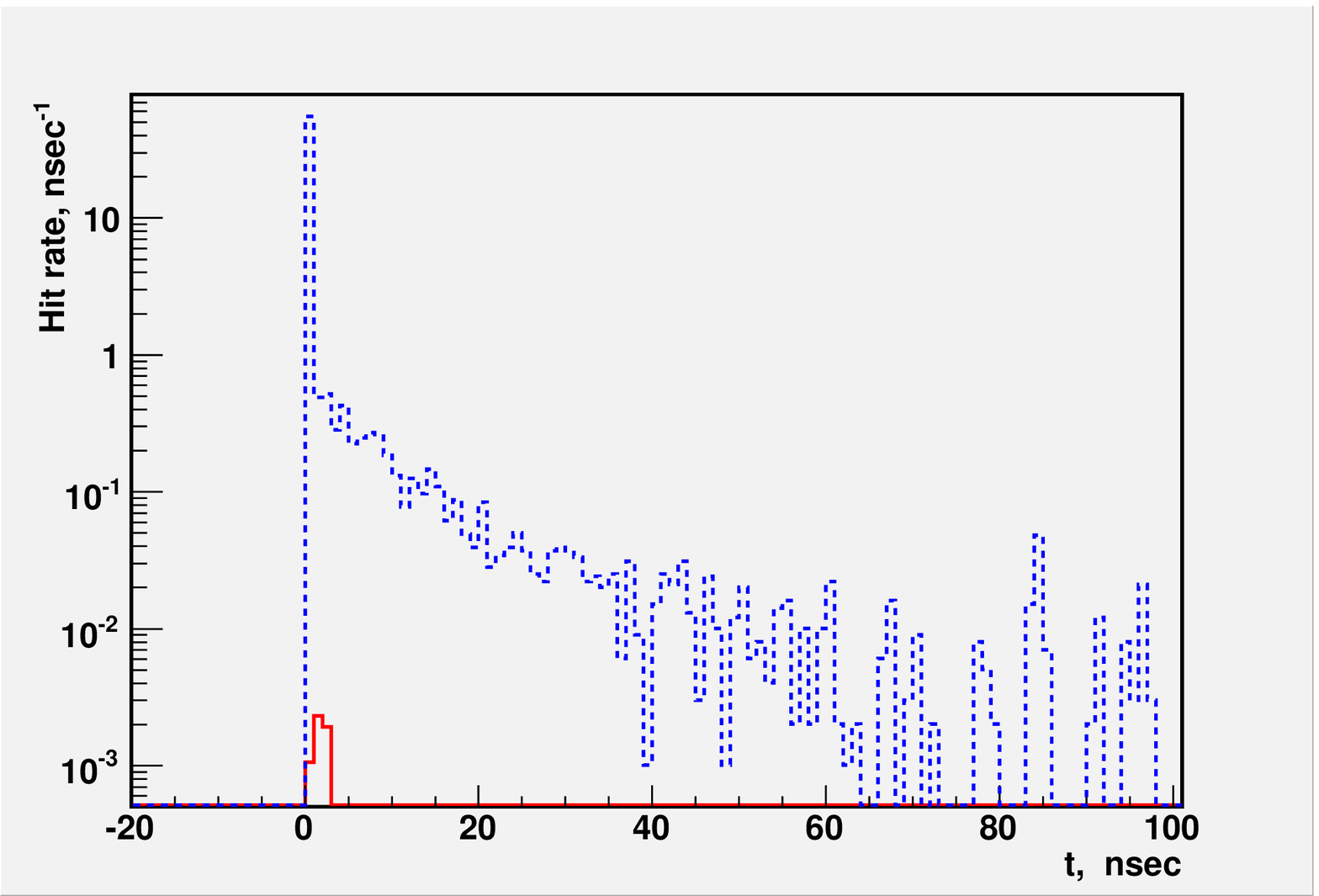,width=\linewidth}
\caption{Time distribution of hit rates in Vertex Barrel. 
Solid line - BDS background (no spoilers), dashed line - $e^+ e^-$ events.
BDS background is from positron tunnel only. }
\end{minipage}
\begin{minipage}[b]{0.91\linewidth}
\centering\epsfig{figure=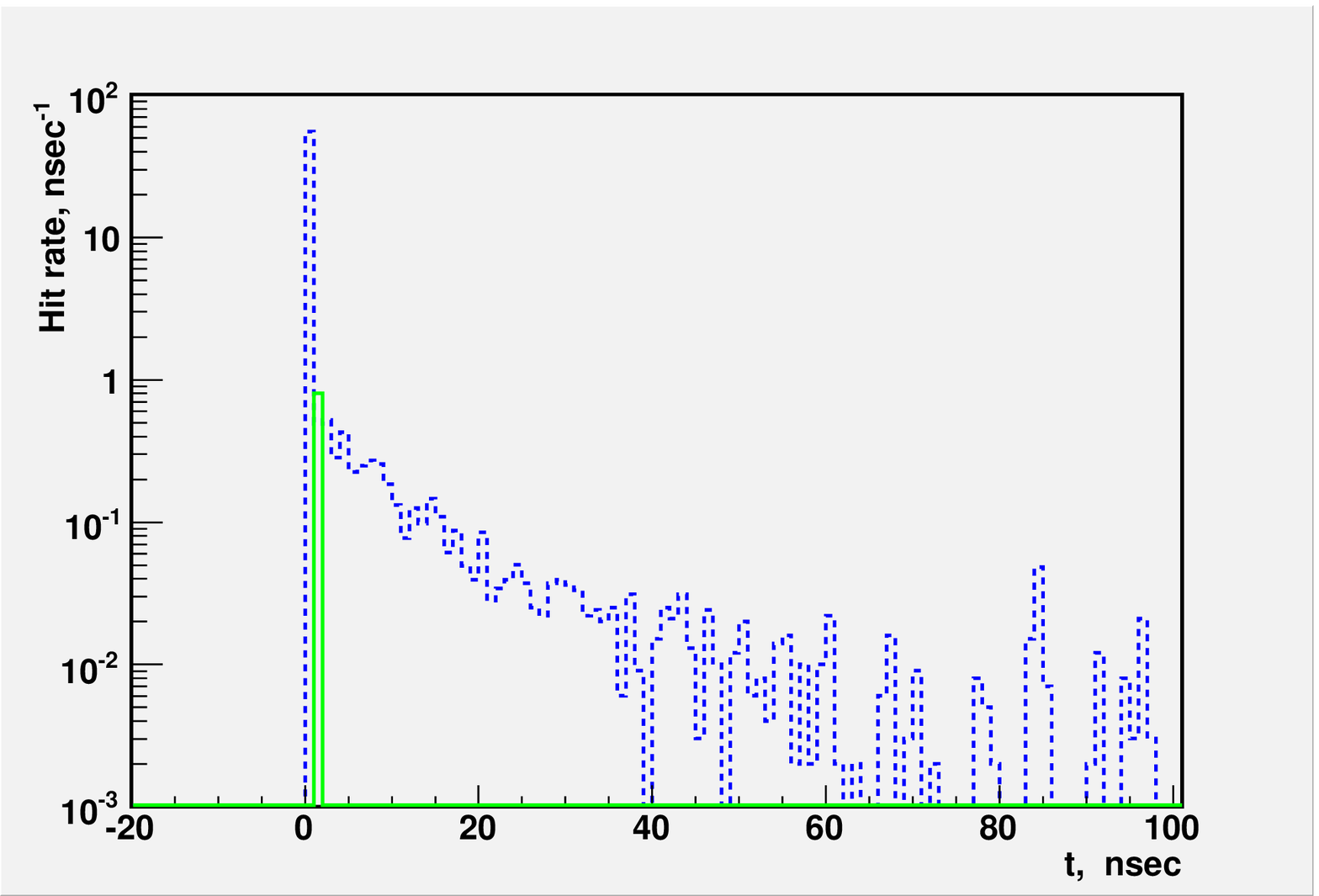,width=\linewidth}
\caption{Time distribution of hit rates in Vertex Barrel. 
Solid line - BDS background (with spoilers), dashed line - $e^+ e^-$ events.
BDS background is from positron tunnel only. }
\end{minipage}
\end{figure}

\newpage
\begin{figure}[hbt!]
\begin{minipage}[b]{0.91\linewidth}
\centering\epsfig{figure=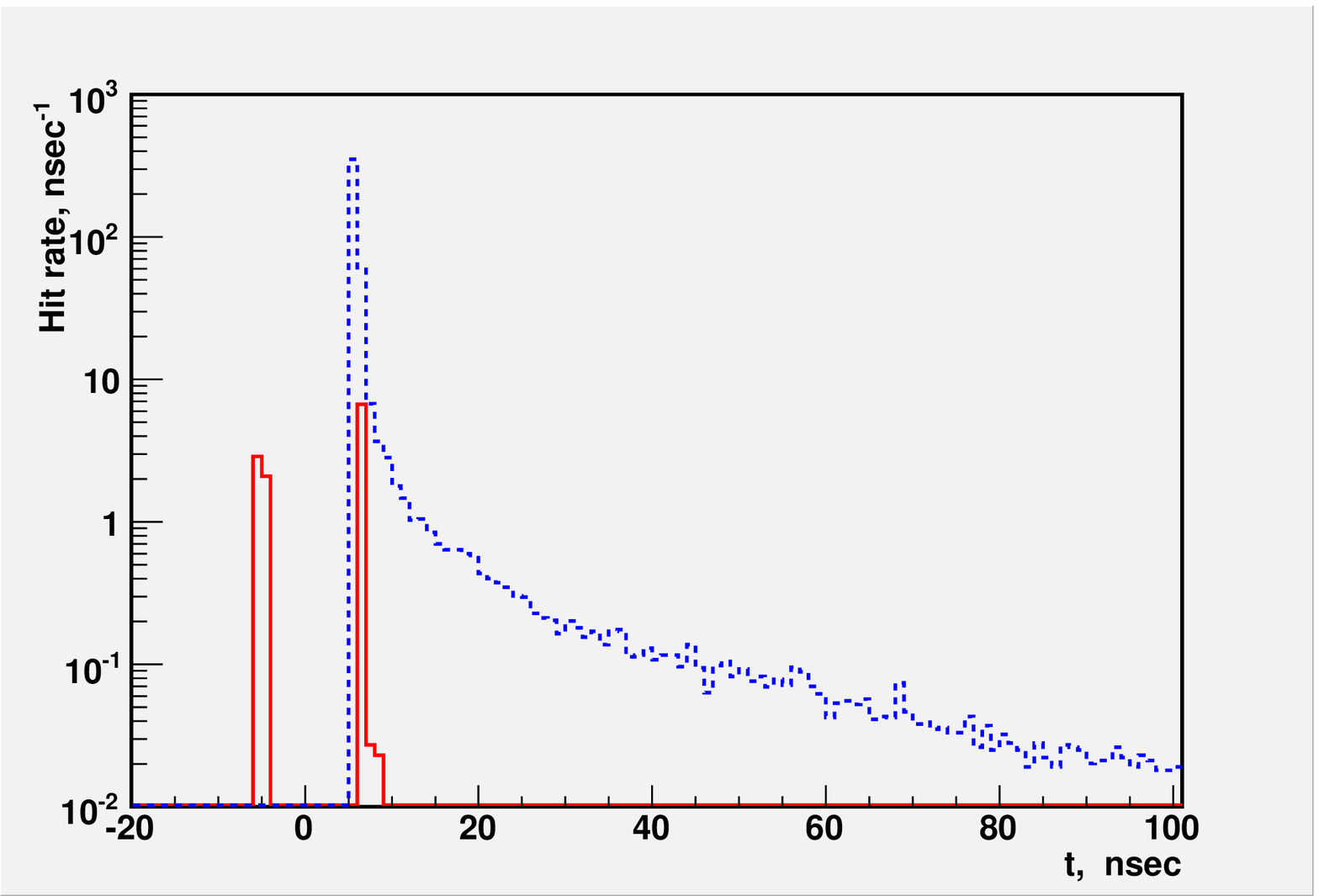,width=\linewidth}
\caption{Time distribution of hit rates in Forward Ecal Endcap. 
Solid line - BDS background (no spoilers), dashed line - $e^+ e^-$ events.
BDS background is from positron tunnel only. }
\end{minipage}
\begin{minipage}[b]{0.91\linewidth}
\centering\epsfig{figure=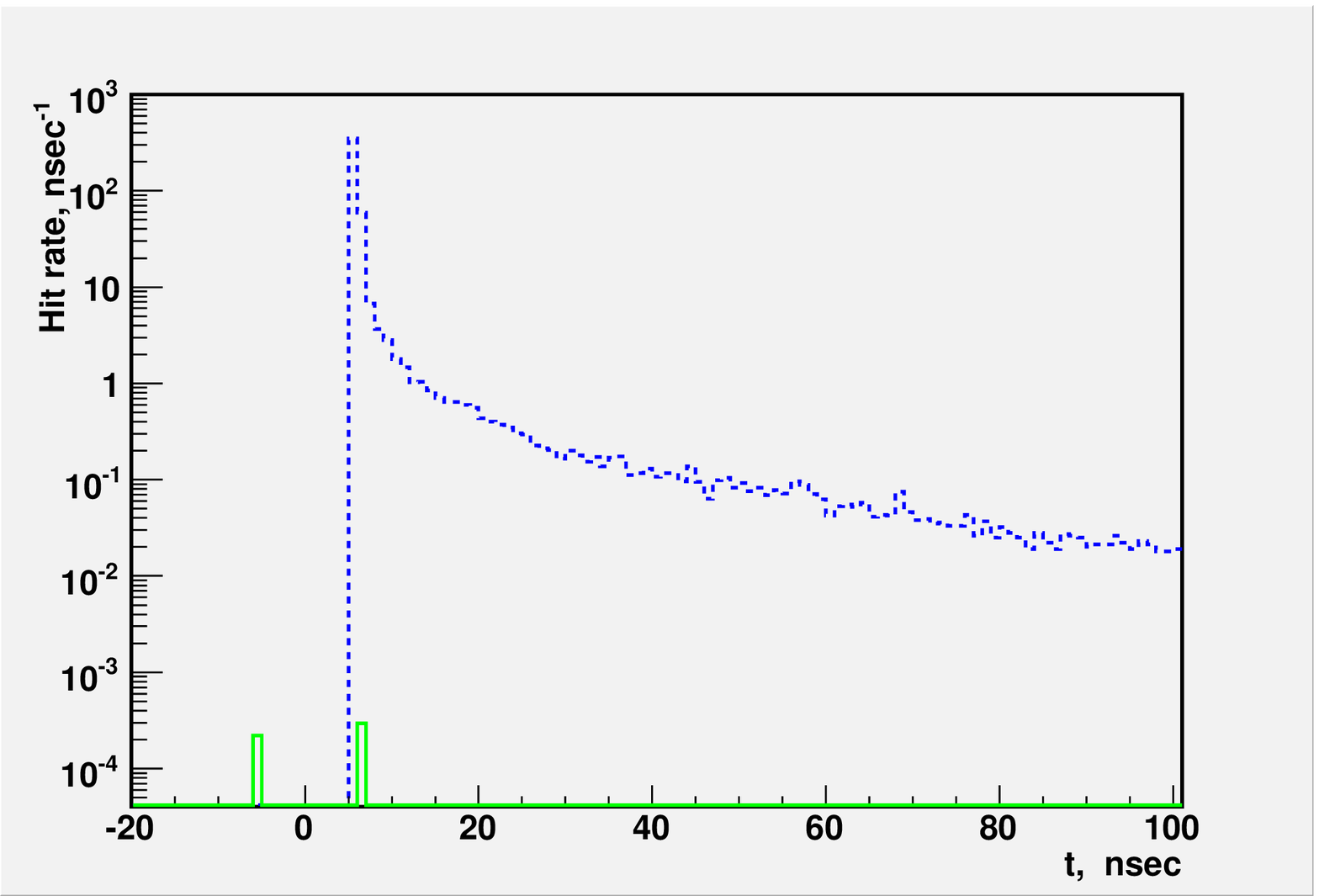,width=\linewidth}
\caption{Time distribution of hit rates in Forward Ecal Endcap. 
Solid line - BDS background (with spoilers), dashed line - $e^+ e^-$ events.
BDS background is from positron tunnel only. }
\end{minipage}
\end{figure}

\newpage
\begin{figure}[hbt!]
\begin{minipage}[b]{0.91\linewidth}
\centering\epsfig{figure=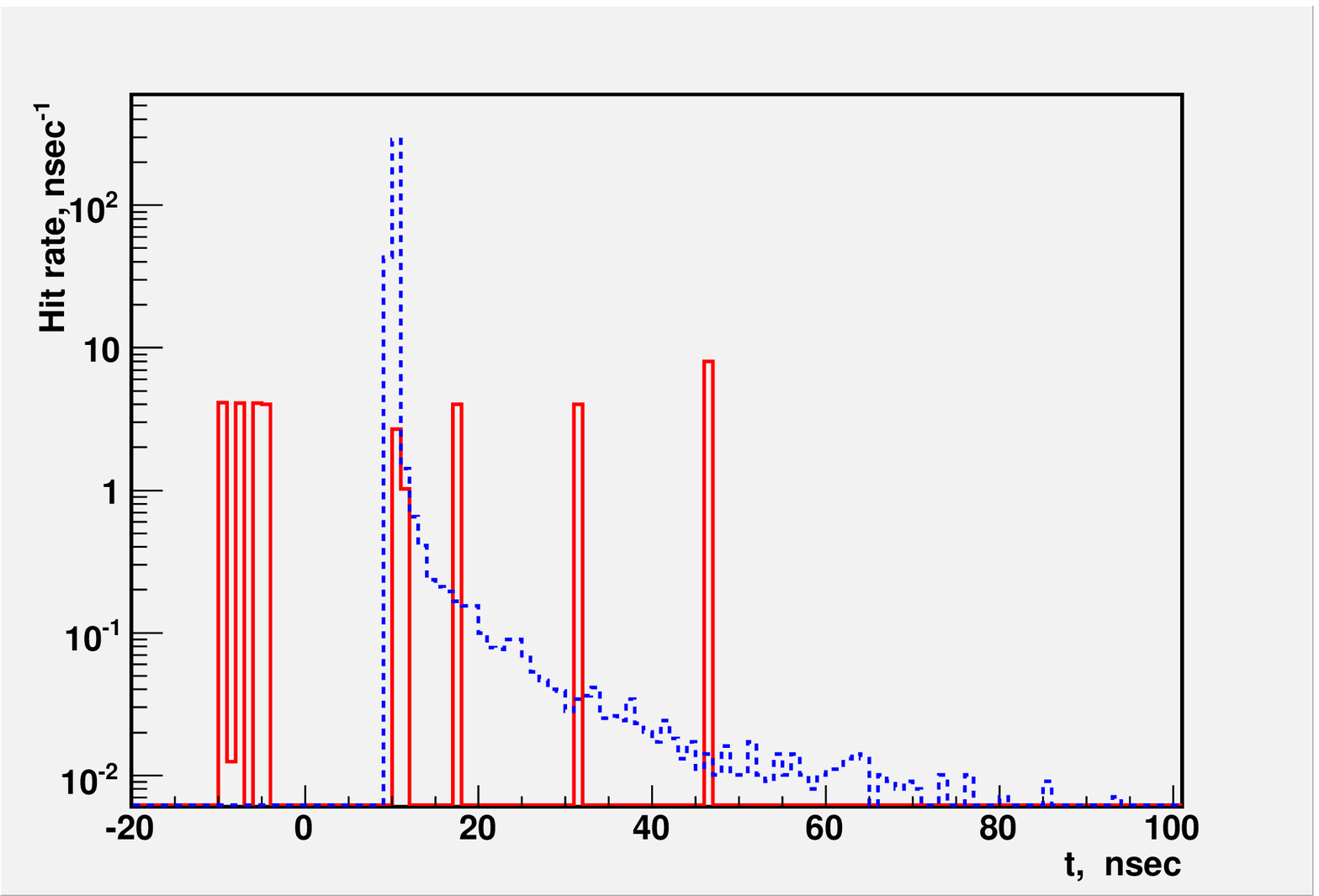,width=\linewidth}
\caption{Time distribution of hit rates in Luminosity Monitor.
Solid line - BDS background (no spoilers), dashed line - $e^+ e^-$ events.
BDS background is from positron tunnel only. }
\end{minipage}
\begin{minipage}[b]{0.91\linewidth}
\centering\epsfig{figure=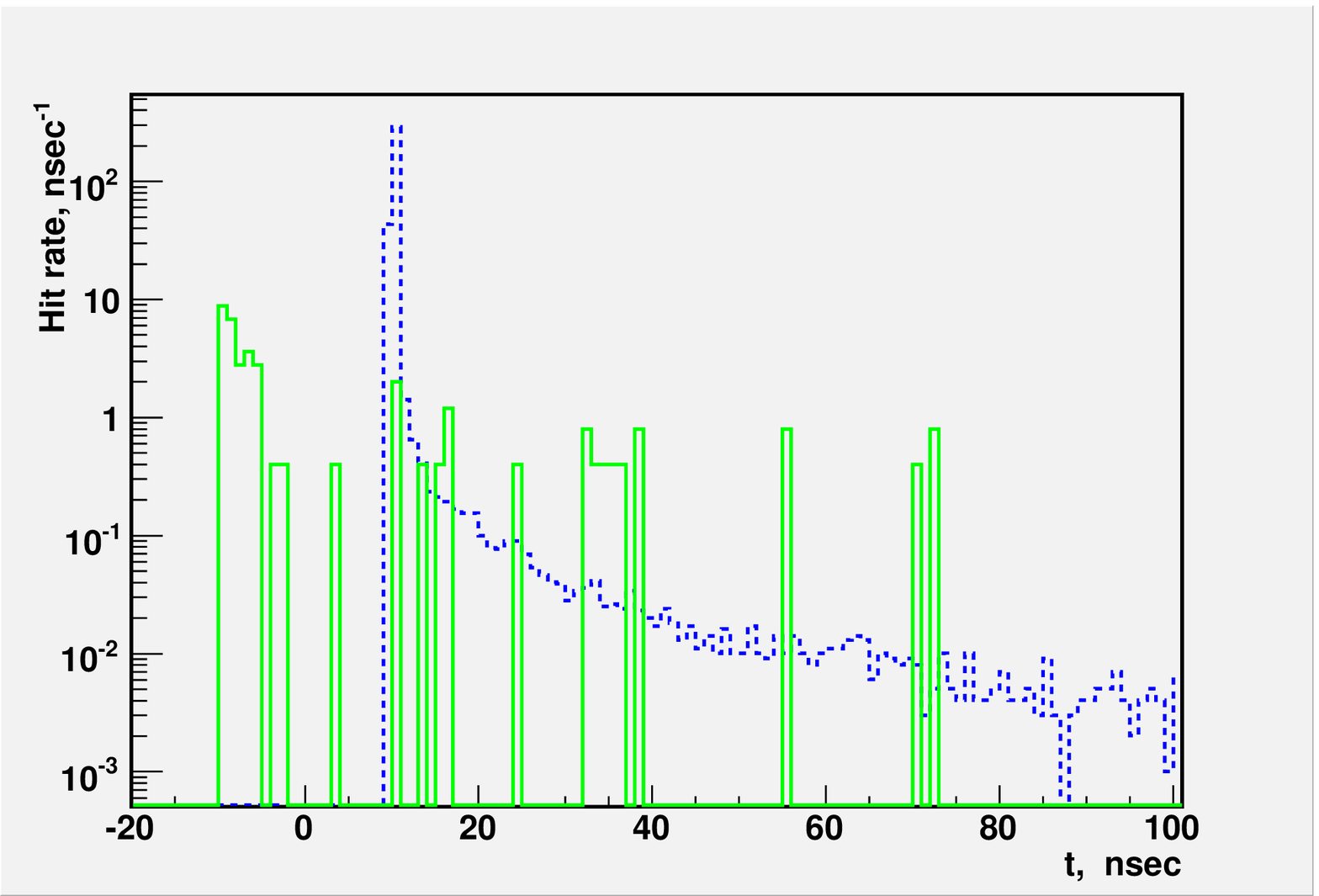,width=\linewidth}
\caption{Time distribution of hit rates in Luminosity Monitor. 
Solid line - BDS background (with spoilers), dashed line - $e^+ e^-$ events.
BDS background is from positron tunnel only. }
\end{minipage}
\end{figure}

\end{document}

%% file: fnaldocheader.tex
\newcommand{\fnaldocheader}[2]{

}